\documentclass[aps,prd,showpacs,superscriptaddress,nofootinbib,preprintnumbers, twocolumn]{revtex4-2}

\usepackage{graphicx,amsmath,amssymb,verbatim,color}
\usepackage{booktabs}
\usepackage[utf8]{inputenc}
\usepackage{fancyhdr, fancybox, empheq} 
\usepackage{color}
\usepackage{amssymb}
\usepackage{breqn}
\usepackage{amsmath}
\usepackage{mathrsfs}
\usepackage{appendix}
\usepackage{amssymb}
\usepackage{float} 

\usepackage[colorlinks=true,citecolor=blue,linkcolor=blue,urlcolor=blue, backref=false,pdfborder={0 0 0}]{hyperref}

\newcommand{\beq}{\begin{equation}}
\newcommand{\eeq}{\end{equation}}
\newcommand{\barr}{\begin{eqnarray}}
\newcommand{\earr}{\end{eqnarray}}

\begingroup\makeatletter
\catcode`,=\active
\global\let\breqn@comma,
\protected\gdef,{\ifmmode\expandafter\breqn@comma\else\expandafter\active@comma\fi}
\endgroup

\begin{document}

\title{Probing Picohertz Gravitational Waves with Pulsars}
 
 \author{Qinyuan Zheng}
\email{qinyuan.zheng@yale.edu}
\affiliation{Department of Physics, Yale University, New Haven, Connecticut 06520, USA}

\author{Chiara M. F. Mingarelli}
\affiliation{Department of Physics, Yale University, New Haven, Connecticut 06520, USA}

\author{William DeRocco}
\affiliation{Maryland Center for Fundamental Physics, University of Maryland, College Park, 4296 Stadium Drive, College Park, MD 20742, USA}
\affiliation{Department of Physics \& Astronomy, The Johns Hopkins University, 3400 N. Charles Street, Baltimore, MD 21218, USA}

\author{Jonathan Nay}
\affiliation{Texas Center for Cosmology and Astroparticle Physics, Weinberg Institute, Department of Physics, The University of Texas at Austin, Austin, TX 78712, USA}

\author{Kimberly K.~Boddy}
\affiliation{Texas Center for Cosmology and Astroparticle Physics, Weinberg Institute, Department of Physics, The University of Texas at Austin, Austin, TX 78712, USA}

\author{Jeff A. Dror}
\affiliation{Institute for Fundamental Theory, Physics Department, University of Florida, Gainesville, FL 32611, USA}

\begin{abstract}
    With periods much longer than the duration of current pulsar timing surveys, gravitational waves in the picohertz (pHz) regime are not detectable in the typical analysis framework for pulsar timing data. However, signatures of these low-frequency signals persist in the slow variation of pulsar timing parameters. In this work, we present the results of the first Bayesian search for continuous pHz gravitational waves using the drift of two sensitive pulsar timing parameters---time derivative of pulsar binary orbital period $\dot{P}_b$ and second order time derivative of pulsar spin period $\ddot{P}$. We apply our new technique to a dataset with more than double the number of pulsars as previous searches in this frequency band, achieving an order-of-magnitude sensitivity improvement. No continuous wave signal is detected in current data; however, we show that future observations by the Square Kilometre Array will provide significantly improved sensitivity and the opportunity to observe continuous pHz signals, including the early stages of supermassive black hole mergers. We explore the detection prospects for this signal by extending existing population models into the pHz regime, finding that future observations will probe phenomenologically-interesting parameter space. Our new Bayesian technique and leading sensitivity in this frequency domain paves the way for new discoveries in both black hole astrophysics and the search for new physics in the early universe.
\end{abstract}

\maketitle

\section{Introduction}

Gravitational waves (GW) have opened an entirely new observational window on the Universe.~\cite{maggiore2008gravitational, afzal2023nanograv, 2016gravitational} Over the past decade, multiple detection techniques have dramatically advanced the field~\cite{sathyaprakash2021gravitational}. Ground-based interferometers such LIGO, Virgo, and KAGRA operate in the $\text{Hz}$ to $\text{kHz}$ range, successfully detecting hundreds of stellar-mass compact binary coalescences~\cite{Abadie_2010}. Space-based interferometry missions, notably LISA~\cite{Amaro-Seoane2017}, are designed to explore the mHz regime. Pulsar timing arrays (PTAs)~\cite{Antoniadis2022, Chen2021, Goncharov2021, nanograv} operate within the nHz regime and have recently uncovered evidence for the stochastic gravitational wave background (GWB) likely produced by supermassive black hole binaries (SMBHB), and potentially other cosmological sources. At the lowest frequencies, between $10^{-18}$ and $10^{-16}$ Hz, the imprints of primordial GWs can be inferred from polarization measurements in the cosmic microwave background (CMB)~\cite{Planck2018Inflation}.

This leaves a significant gap in GW frequency band between $10^{-16}$ Hz and 1 nHz, called the picohertz (pHz) regime, where exciting physics may be hiding. First of all, here lies a population of pHz frequency SMBHBs -- progenitors of the nHz SMBHB population found in the PTA band~\cite{Begelman1980}. All these binaries emit continuous GWs (CW), and a GWB arises from the superposition of all these individual CW sources. In the pHz regime, environmental effects likely dominate over GW emission in the binary evolution, which should influence the SMBHB demographics. Detecting GWs in this pHz frequency regime can teach us about the SMBHB population, and in turn give us insight into the physics driving SMBHB evolution. 

A range of beyond-the-Standard-Model and early-Universe processes could also contribute to the GW signals in this band. These include networks of cosmic strings~\cite{Ghoshal2023, Chang2020}, first-order phase transitions associated with symmetry breaking events in the early Universe~\cite{Freese2023, Weir2018}, turbulent processes during the quantum chromodynamics epoch~\cite{Neronov2021, Brandenburg2021, Moore2021}, and primordial GWs generated during inflation~\cite{Liddle2000, Kamionkowski2016} and subsequent preheating phases~\cite{Boyle2005}. Detection of pHz GWs would thus open a new observational frontier for testing new physics.

\begin{figure*}[ht!]
\centering
\includegraphics[width=0.75\linewidth]{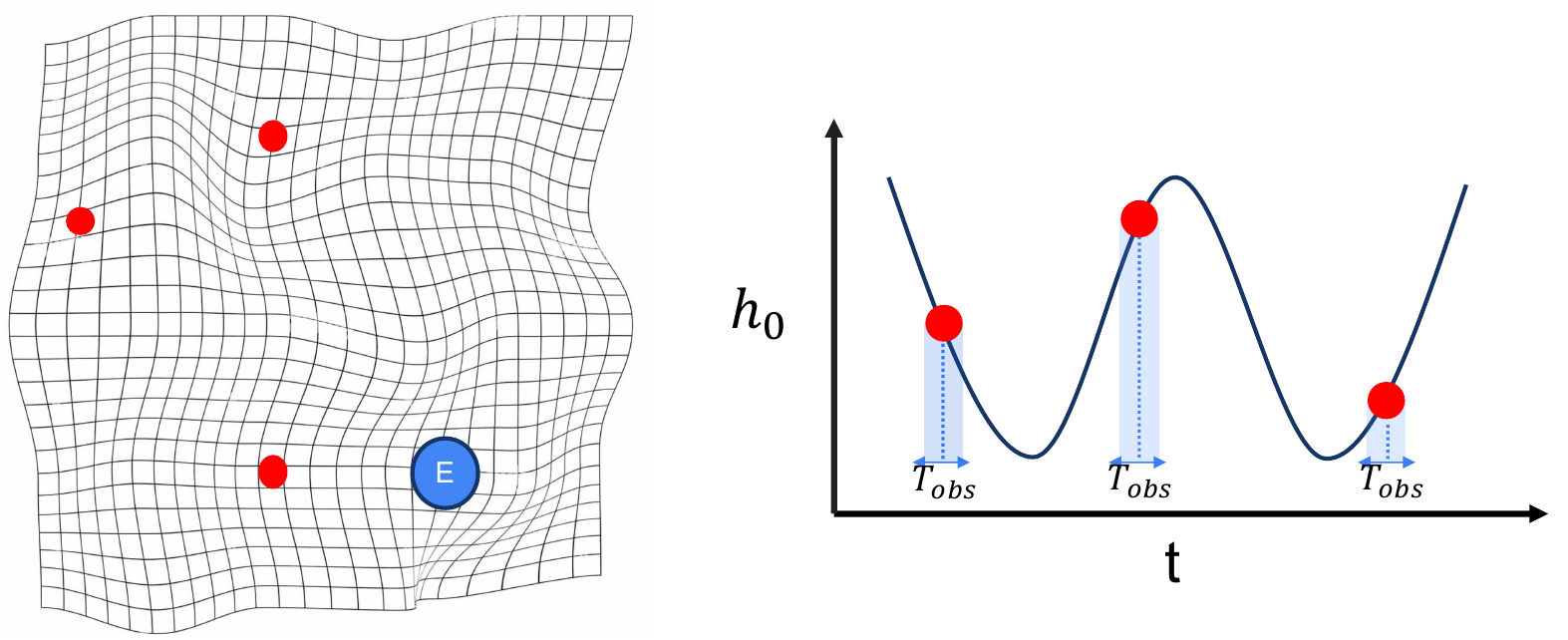}
\caption{Pulsar timing parameters capture the metric perturbations due to pHz GWs. LHS: the pulsars (red) and the Earth (blue) are both affected by the passing GW. RHS: short observation baselines ($T_{\mathrm{obs}}$) allow for treating the perturbations as stationary in time. Pulsars at different locations are probing the CWs emitted by the source at different times. For a 10 pHz CW, the period is $\sim3,000$ years, while the observation time is $\sim10$ years. The metric perturbations manifest as measurable redshift $v_{\mathrm{GW}}$ and the associated derivatives $a_{\mathrm{GW}}$ (acceleration) and $j_{\mathrm{GW}}$ (jerk). The explicit expressions of these quantities in terms of the CW waveform are given in Appendix~\ref{sec:analytic derivation}.}
\label{fig:illustration}
\end{figure*}

Traditional PTAs are sensitive to nHz GWs. To probe the pHz frequency regime, we use millisecond pulsars (MSPs) in a novel way: instead of analyzing the timing residuals, we search for drifts in MSP parameters induced by ultra-low frequency GWs. In this pHz frequency range, GWs have periods of hundreds to thousands of years. For observation times of only a few decades, we capture just a small segment of the gravitational waveform. These pHz GWs induce distortions in the spacetime around the pulsars that manifest as secular changes in the observed pulsar parameters, e.g. $\dot{P}_b,~\ddot{P}$. $\dot{P}_b$ is the time derivative of the orbital period of a pulsar with a binary stellar companion, while $\ddot{P}$ is the second-order time derivative of the pulsar spin period.

The extensive data accumulated by PTA campaigns, combined with increasingly sophisticated pulsar noise models~\cite{Goncharov2024, Larsen2024, Agazie2023}, provide an unprecedented opportunity to explore the pHz frequency regime. The concept that pHz GWs can induce measurable perturbations in pulsar timing parameters was first proposed by~\citet{Bertotti1983}, and later expanded upon by~\citet{DeRocco2023, DeRoccoDror2024} in a modern theoretical framework. Here we present a new Bayesian framework that uses MSP timing parameters to search for CWs in the pHz frequency band. We develop an analytic model that describes the CW-induced drifts in the observed pulsar parameters $\dot{P}_b$ and $\ddot{P}$. 

For our study, we extend an existing GWB model for SMBHBs~\cite{Casey-Clyde2022} to pHz frequencies. The population is matched in a self-consistent way to the evidence for a GWB in the NANOGrav 15-year data~\cite{nanograv}. From this extended model, we derive the characteristic strain of the background and generate synthetic realizations of pHz CW sources. Our analysis improves the upper limits on CW strain amplitudes by up to an order of magnitude relative to~\citet{DeRoccoDror2024}, and achieves a new leading constraint. Furthermore, through detailed simulations, we explore the CW scaling laws in the pHz regime and forecast strain amplitude sensitivities accessible by next-generation telescopes such as the Square Kilometre Array (SKA). We show that by timing new pulsars and improving timing precision with SKA, the technique presented here could potentially detect individual pHz sources, opening a new avenue to probe fundamental physics not accessible by any other means.

The paper is organized as follows. In Section~\ref{sec:method}, we introduce the theoretical basis for the pHz GW detection method and the Bayesian framework. Section~\ref{sec:SMBHB} describes the construction and extension of the SMBHB population model into the sub-nHz regime. Section~\ref{sec:dataset} presents the datasets and processing techniques employed in this analysis. In Section~\ref{sec:result}, we summarize the results of our search. Section~\ref{sec:discuss} discusses the implications of our findings and prospects for future work.

\section{Methods}
\label{sec:method}
\subsection{Analytical Model}

In traditional PTA analyses, the lowest detectable GW frequency is determined by the inverse of the observation time, with $f_{\mathrm{GW}}^{\mathrm{min}} \geq T_{\mathrm{obs}}^{-1}$. Potential GW signals at frequencies below $f_{\mathrm{GW}}^{\mathrm{min}}$ are absorbed into the timing model noise during the fitting procedure. Nevertheless, these pHz GWs still induce systematic shifts in the measured pulsar parameters. 

Specifically, at pHz frequencies, a passing GW induces an additional effective velocity between the pulsar and the Solar System Barycenter (SSB). The induced velocity is given by
\begin{equation}
v_{\mathrm{GW}}=\sum_{A=+,\times}F_A(\hat{\Omega})\left[h_A(t,0)-h_A(t-d_p,\bold{d}_p)\right] \, ,
\label{eq:1}
\end{equation}

where the subscript $A=+,\times$ labels the GW polarizations (plus and cross, respectively), and $\bold{d}_p$ denotes the vector pointing from the SSB to pulsar $a$. The functions $h_{+,\times}(t,\bold{x})$ represent the GW strain at position $\bold{x}$ and time $t$. The antenna beam pattern, $F_A(\hat{\Omega})$, is defined as~\cite{Romano_2024}
\begin{equation}
\label{eq:antenna}
F_A(\hat{\Omega})=\frac{\hat{d_p^i}\hat{d_p^j}\hat{e}_{ij}^A(\hat{\Omega})}{2(1+\hat{\Omega}\cdot \hat{d_p})}\, ,
\end{equation}

where $\hat{e}_{ij}^A$ is the polarization tensor of the GW, and $\hat{\Omega}$ is the unit vector of GW propagation. We work in the transverse traceless gauge and $i,j$ are the spatial indices.

Here we focus on the CW search using $\dot{P}_b$ and $\ddot{P}$. At sufficiently low GW frequencies, the additional velocity term induced by the GW also creates an apparent acceleration. This manifests as a constant offset in the pulsar timing parameters over the observation time. See Fig.~\ref{fig:illustration} for a pictorial illustration.

To leading order, the observed derivative of the binary orbital period can be written as~\cite{Bell1996}

\begin{equation}
\label{eq:Pbdot_measurement}
\frac{\dot{P}_{b, \mathrm{obs}}}{P_b}=\frac{\dot{P}_{b, \text{int}}}{P_b}-a_{\text{shk}}-a_{\text{MW}}-a_{\text{GW}}\, ,
\end{equation}
where the terms on the right hand side correspond to, in order: $\dot{P}_{b, \text{int}}/{P_b}$, the intrinsic period derivative due to GW emission from the binary; $a_{\text{shk}}$, the kinematic Shklovskii effect due to the motion of the pulsar proportional to the relative motion of the pulsar perpendicular to the line of sight; $a_{\text{MW}}$, the acceleration due to the Milky Way gravitational potential; and $a_{\text{GW}}$, the contribution from pHz frequency GWs, which constitutes the signal we seek. In particular, $a_{\text{skh}}=\mu^2 d_p$, where $\mu$ is the proper motion of the pulsar.

Similarly, the observed second derivative of the spin period is related to the jerk induced by GWs to leading order through
\begin{equation}
\label{eq:pddcalc}
\frac{\ddot{P}_{\mathrm{obs}}}{P}=j_{\text{GW}}\, .
\end{equation}

In principle, there are kinetic and galactic contributions to $\ddot{P}_{\mathrm{obs}}$ as well, but the non-relativistic nature of the galaxy makes them subdominant to $j_{\mathrm{GW}}$ by several orders of magnitude~\cite{DeRoccoDror2024}. Intrinsic contributions due to magnetic dipole braking are of order $(\dot{P}/P)^2$, which typically evaluate to $\sim 10^{-35} s^{-2}$, much below the observed uncertainty of $\ddot{P}_{\mathrm{obs}}/P$~\cite{LorimerKramer2012}. The higher order parameters such as $\dddot{P}$ can be derived in a similar fashion, and could be used for GW search in the future analysis. Explicit expressions for $a_{\mathrm{GW}}$ and $j_{\mathrm{GW}}$ induced by a CW are derived in Appendix~\ref{sec:analytic derivation}.

\subsection{Bayesian Analysis}

For this study, we consider GWs emitted by SMBHBs in a circular orbit. However, the method is easily generalizable to other CW sources. We have used 7 parameters characterizing the CW signal: $h_0$, the GW strain amplitude; $f_{\text{GW}}$, the GW frequency; and $\bold{\Theta}$, which consist of $\{\alpha, \delta, i, \phi, \Psi_0\}$.  In order $\bold{\Theta}$ include the right ascension and declination $\alpha, \delta$, the orientation of the GW source $i, \phi$, and the orbital phase of the SMBH binary $\Psi_0$. Due to the relatively high dimensionality of the problem, we adopt a Bayesian approach. 

We construct a likelihood making the standard assumption that pulsar uncertainties are uncorrelated~\cite{LorimerKramer2012}. The likelihood is given by
\begin{widetext}
\begin{equation}
    \mathscr{L}(h_0, f_{\mathrm{GW}}, \bold{\Theta}|{\vec{y}_{\mathrm{GW}}})=\prod_{p=1}^{N}\frac{1}{\sqrt{2\pi}\sigma_p}\exp\left[-\frac{(y_{\mathrm{GW},p}-\bar{y}_{\mathrm{GW},p}(h_0, f_{\mathrm{GW}}, \bold{\Theta}))^2}{2\sigma_p^2}\right]\, ,
\end{equation}
\end{widetext}
where $\vec{y}_{\mathrm{GW}}$ the collection of accelerations or jerks $y_{\mathrm{GW},p} = \{a_{\mathrm{GW},p}, j_{\mathrm{GW},p}\}$ [calculated using Eq.~\ref{eq:Pbdot_measurement} or~\ref{eq:pddcalc}] with corresponding uncertainty $\sigma_p$ for each pulsar $p$. $N$ is the number of pulsars in the data.

Bayes' theorem states that the posterior distribution of parameters \( \theta \) given data \( D \) is:

\begin{equation}
    P(\theta | D) = \frac{P(D | \theta) P(\theta)}{P(D)}\, ,
\end{equation}
where:
\( P(\theta | D) \) is the posterior distribution,
\( P(D | \theta) \) is the likelihood function,
\( P(\theta) \) is the prior distribution,
and the denominator is the evidence.

In this study, we adopt uniform priors for all parameters, as listed in Table~\ref{table:priors}. However, we note that informative priors could apply if, for example, MMA observations are available. Particular care should be given to the strain amplitude $h_0$. By the nature of a GW search where the signal is weak, $h_0$ resides in a linear space spanning multiple orders of magnitude. When looking for a GW signal, we use the logarithmic uniform prior so that smaller values of $h_0$ are sampled sufficiently. In the case where the data are not sufficiently informative so that the likelihood function goes flat for small values of $h_0$, we use the linear-uniform prior, which corresponds to a prior in the log space where small values of $h_0$ are exponentially suppressed. It is worth noting that if the data are informative enough, the prior should not influence the posterior distribution after convergence. However, when the data are too noisy to support a detection, the linear flat prior gives a conservative estimate of the upper limit of $h_0$ in the sense that false positives tend to be avoided. The full prior is the product of the individual priors:

\begin{equation}
    P(\theta)=\pi(\alpha)\pi(\delta)\pi(i)\pi(\phi)\pi(\Psi_0)\pi(f_{\text{GW}})\pi(h_0)\, .
\end{equation}

\renewcommand{\arraystretch}{2}

\begin{table}[h]
\centering
\begin{tabular}{lll}
\toprule
\textbf{Parameter} & \textbf{Prior Distribution} & ~~~\textbf{Domain} \\
\midrule
$\alpha$       & uniform               & $0^\circ \leq \alpha < 360^\circ$ \\
$\delta$       & $\displaystyle \frac{1}{2} \cos(b)$     & $-90^\circ \leq \delta < 90^\circ$ \\
$i$            & $\displaystyle \frac{1}{2} \cos(i)$           & $-\frac{\pi}{2} < i < \frac{\pi}{2}$ \\
$\phi$         & uniform               & $-\frac{\pi}{2} < \phi < \frac{3\pi}{2}$ \\
$\Psi_0$       & uniform               & $0 < \Psi_0 < 2\pi$ \\
$\log_{10}f_{\mathrm{GW}}$            & uniform                 & $-12 < f < -9$ \\
$\log_{10}h_0$            & uniform                                          & $-20 < h_0 < 0$ \\
\bottomrule
\end{tabular}
\caption{Prior distributions for model parameters.}
\label{table:priors}
\end{table}

We run Markov Chain Monte Carlo (MCMC) to determine the posterior distribution and interpret the posterior for either detection of a signal or setting an upper bound using existing data. Our method is naturally poised for using more informative priors based on multi-messenger astronomy observations, the details of which are beyond the scope of this paper. Such informative priors will help break the degeneracies and pick out the signal more accurately.

To characterize a detection, we employ a combined analysis based on the Bayes factor (BF) and the highest posterior density region. Specifically, the marginalized posterior distribution for the strain amplitude must be clearly distinct from the null hypothesis to declare a detection. We compute the BFs using the Savage-Dickey density ratio method. To set a detection threshold, we simulate the null distribution of the BF for the data under analysis.

If the marginalized posterior for the strain amplitude were to support a CW detection, we could infer the other relevant source parameters by analyzing the high density regions of the posterior distributions. Since the current data do not support a CW detection, we defer detailed parameter estimation to future analyses.

\section{Supermassive Black Hole Binary Population Model}
\label{sec:SMBHB}

In order to understand the detection prospects of our method, we extend the SMBHB population synthesis model from \cite{Casey-Clyde2022} to the pHz regime. This model takes the GWB amplitude measurement as an input, and uses it to constrain SMBHB population characteristics, including the minimum black hole mass contributing to the background, the number density of SMBHBs per unit volume, and the volume that encloses the GWB. Here we use this population synthesis code to anchor the pHz SMBHB population to the nHz population, creating a self-consistent model of our GW sources. 

In this model, we use dual active galactic nuclei (AGN) as tracers of SMBHBs: we assume a fixed ratio between dual AGN density $\Phi_{\mathrm{Q}}$ and SMBHB density $\Phi_{\mathrm{SMBHB}}$, depending on a mass ratio distribution $p(q)$:

\begin{equation}
\frac{d^3 \Phi_{\mathrm{SMBHB}}}{dM_{\mathrm{BH,1}}\, dz\, dq} = R \frac{d^2 \Phi_{\mathrm{Q}}}{dM_{\mathrm{BH}}\, dz} \,p(q)\, ,
\end{equation}
where $M_{\mathrm{BH,1}}$ is the mass of the primary, $M_{\mathrm{BH}}$ is the total mass of the binary, $z$ is the redshift, and $q$ is the mass ratio. The model parameters~\cite{Casey-Clyde2022} including the constant of proportionality $R$ are then extracted by fitting the integrated GWB characteristic strain to the evidence of GWB in the NANOGrav 15-year data. 

Towards the pHz regime, several effects that shrink the SMBHB orbits are likely to compete with or even dominate over the GW emission of the binary system. The dual--AGN that contain the SMBHs which will eventually comprise the binary initially evolves under dynamical friction~\cite{BinneyTremaine2008}. As the SMBHs approach a separation of $\sim100$~pc, the evolution of the dual--AGN becomes dominated by three-body interactions with individual stars, also known as ``stellar hardening". At separations of $\sim1$ pc stellar hardening becomes inefficient as the SMBHs clears their vicinity of nearby stars. At this stage interactions between the SMBHB and nearby gas may become important for further shrinking the SMBH separation, though the details of this process remain an open question. Hard binaries of SMBHs start to form at this separation. For a SMBHB of mass $10^{8}\sim 10^{10}\mathrm{{M}}_{{\odot}}$ and separation $\sim 1~\text{pc}$, the emitted GW frequencies are $\sim 10-100~\text{pHz}$.

In our model, we include both dynamical friction and stellar hardening effects. We do not include the gas hardening effect and leave it for future investigation. Our method of pHz GW detection would be an important step to probe the various models of SMBHB evolution, manifested as a turnover that deviates from the power-law spectrum of GWB in the PTA frequency band. The GWB from our population model is orders of magnitude below the CW upper limits, so the noise due to GWB can be reasonably ignore in this work.

\section{Dataset}
\label{sec:dataset}

For the $\dot{P}_b$ analysis, we use the updated data from~\cite{Moran:2023}, which includes 29 pulsars. For our purposes, we independently estimate the Galactic accelerations of the pulsars using the MWPotential2014 model as implemented in the \texttt{galpy} package~\cite{Bovy2015}. 

The original data set from~\cite{Chakrabarti_2021} used in~\citet{DeRoccoDror2024} contains 14 pulsars, 13 of which are also included in our updated sample. After cross-checking with other recent data releases, we adopt the more updated measurements from~\cite{Moran:2023} for all overlapping pulsars. In total, we obtain 30 pulsars with updated timing measurements, which is double the number used in the analysis of~\citet{DeRoccoDror2024}. The observed $\dot{P}_b$'s range in order of magnitude from $10^{-21}$ to $10^{-8}$, with the majority being around $10^{-14}\sim 10^{-12}$. We subtract non-GW contributions from the observed $\dot{P}_b$ and report the GW induced $\Delta \dot{P}_b/P_b$ data immediately used for analysis. The timing baselines of the pulsars range from 4 years to 22 years, with $\mu s$ level RMS timing residuals. For a full list of the pulsars, see Table~\ref{tab:pbdot_pulsar_parameters}.

\begin{figure*}
    \centering
    \includegraphics[width=0.9\textwidth]{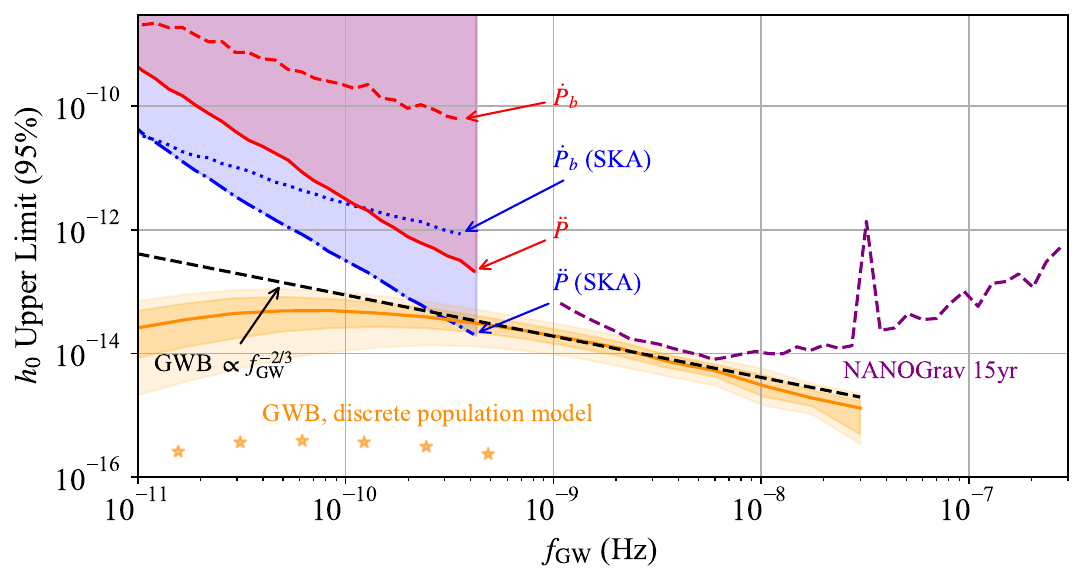}
    \caption{CW strain amplitude upper limits and forecast detection sensitivities of pulsar timing parameters $\dot{P}_b$ and $\ddot{P}$. The fiducial CW sources are modeled as circular orbit SMBHBs. The current upper limit in this work is an order of magnitude better than best constraints from previous work in this frequency range~\cite{DeRoccoDror2024}. Forecast sensitivities with SKA data provide another order of magnitude improvement by increase of data size alone, without accounting for the improvement in data uncertainty. Current results are comparable to PTA upper limit on CWs (purple dashed line) for certain frequencies. Expected individual CW sources (denoted by the stars) are at least 3 orders of magnitude in strain amplitude below the current constraint according to our SMBHB population model. The orange solid line is the GWB spectrum drawn from our population model by matching self-consistently with GWB at $f_{\mathrm{GW}}=1~\mathrm{yr}^{-1}$ detected by NANOGrav. The shaded regions centered on this curve denotes the uncertainty quantiles. The black dashed line represents the power-law GWB spectrum matched to NANOGrav GWB at $f_{\mathrm{GW}}=1~\mathrm{yr}^{-1}$.}
    \label{fig:upbd}
\end{figure*}

Pulsar distance measurements are essential for reducing degeneracies in this analysis. If the resulting dataset is large enough, we require $f\Delta d_p\leq 1$; namely, we constrain the pulsar distance up to a precision less than a radiation wavelength, which helps mitigate the degeneracies associated with $2\pi$ periodicity of trig functions. For the pHz regime, this typically corresponds to a wavelength of $\sim 100$~pc. Half of the pulsars in the $\dot{P}_b$ data meet this criterion and are used in the analysis, as we use $f_{\mathrm{GW}}=100$ pHz to define the cutoff. When we perform our MCMC analysis, we use the pulsar distance measurements as a prior, and sample in the restricted error range in the analysis.

Additionally, to ensure that the Taylor expansions used in the analytic modeling remain valid, we cut off the analysis at $(4T_{\mathrm{max}})^{-1}$, consistent with \cite{DeRoccoDror2024}, where $T_{\text{max}}$ is the longest observation time in the data.

We account for the Shklovskii contribution $a_{\text{kin}}$ using the analytic formula in Section~\ref{sec:method}. The uncertainty in $d$ dominates the uncertainty in $a_{\text{kin}}$, allowing us to treat $\mu$ as effectively fixed with a small uncertainty. When sampling $d$, we will propagate the proper motion uncertainty and add it in quadrature to the total uncertainty in $a_{\text{kin}}$. We verify that our cut on $\dot{P}_b$ data has resulted in a better upper limit by simulating null distributions, finding that including all pulsars shifts the null Bayes factor distribution towards larger values, which in turn biases the posterior toward higher strain amplitudes. The values of $\mu$ used in our analysis are taken from~\cite{Moran:2023}.

For the $\ddot{P}$ analysis, we use the same data as in~\cite{DeRoccoDror2024}, given by~\cite{Desvignes2016} and~\cite{Reardon2016}. A full list can be found in Table~\ref{tab:pdd_pulsar_parameters}.The pulsars have timing baselines ranging from 7 to 18 years, with a typical RMS timing residual of a few $\mu s$. We use the parallax measurement of the pulsar distance whenever it is available. Otherwise, we estimate using dispersion measure (DM) and a model of Galactic free electron density model. A total of 13 pulsar distances in the data are estimated using DM and the Galactic free electron density model YMW 17~\cite{yao2017new}. Since they lack a well-defined uncertainty in distance, we conservatively assume a $40\%$ relative uncertainty on these distances. Only one fourth of the pulsar distances in the $\ddot{P}$ data satisfy $f\Delta d_p\leq 1$, and dropping the majority of the data will make the set prone to outliers and lose information from a larger data set. Thus, we retain all pulsars in the $\ddot{P}$ data for analysis. 

We also include an estimate of the red noise following the procedure described in~\cite{DeRoccoDror2024}. The red noise is treated as a stochastic background given by a power-law spectrum, and we estimate the ensemble variance of the two timing parameters of interest due to the red noise. For the current data, red noise contributions are negligible in the $\dot{P}_b$ measurements but relevant for $\ddot{P}$. We incorporate the red noise contribution into the total uncertainty budget for $\ddot{P}$. We emphasize that as timing precision improves, red noise may no longer remain subdominant to other sources of uncertainty in $\dot{P}_b$ and will need to be treated carefully in future analyses. The full data can be found in Appendix~\ref{sec:pulsar data}.

\section{Results}
\label{sec:result}

\begin{figure}[ht!]
\centering
\includegraphics[width=0.98\columnwidth]{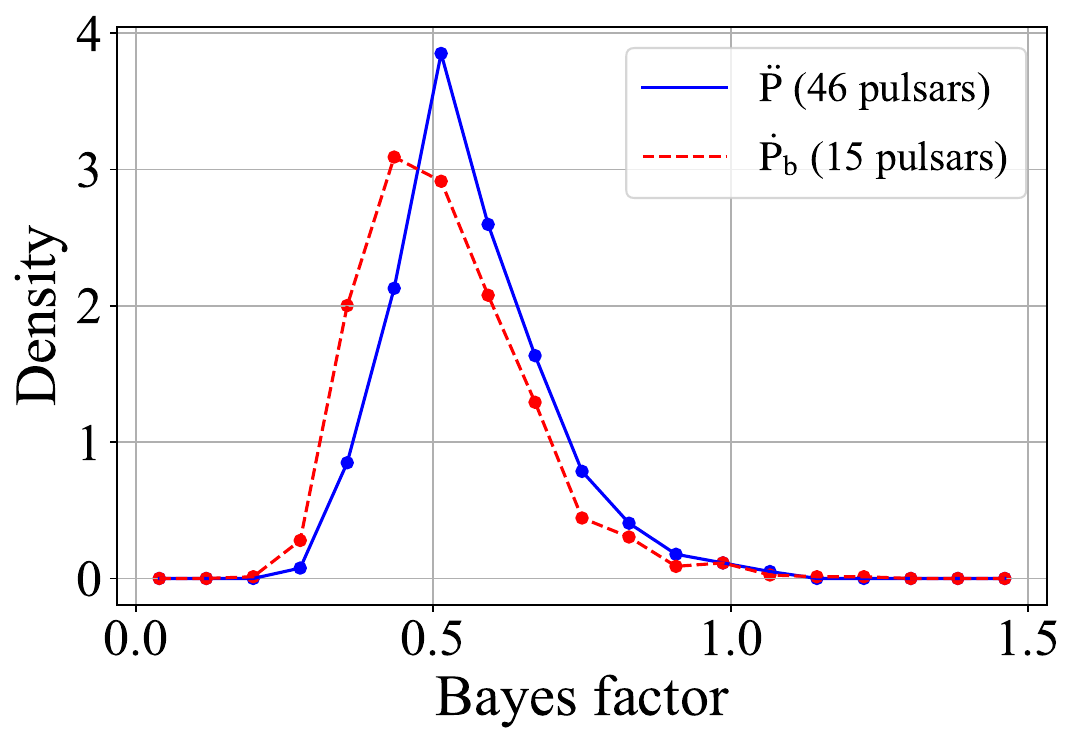}
\caption{Null distributions of the $\dot{P}_b$ and $\ddot{P}$ data used to place current upper bounds on CW strain amplitude. Both have a small tail surpassing BF=1. In each case, a null CW signal is injected to the simulated data, and $1,000$ realizations of BFs are extracted.}
\label{fig:null_dist}
\end{figure}

We search for CW signals using $\dot{P}_b$ and $\ddot{P}$ independently. For the $\dot{P}_b$ data, a total of 15 pulsars satisfy our criteria in Sec.~\ref{sec:dataset}, which we use for GW searches and to place upper limit on the strain amplitude. To determine a detection threshold, we calculate the null distributions for both data sets through simulations with $1,000$ samples each. The null distributions exhibit small tails extending beyond $\text{BF}=1$, with $100\%$ of the samples falling below $\text{BF}=3$. We choose the conservative $\text{BF}=3$ as the detection threshold. (See Fig.~\ref{fig:null_dist}) 

\begin{figure*}[ht!]
    \centering
    \includegraphics[width=1\linewidth]{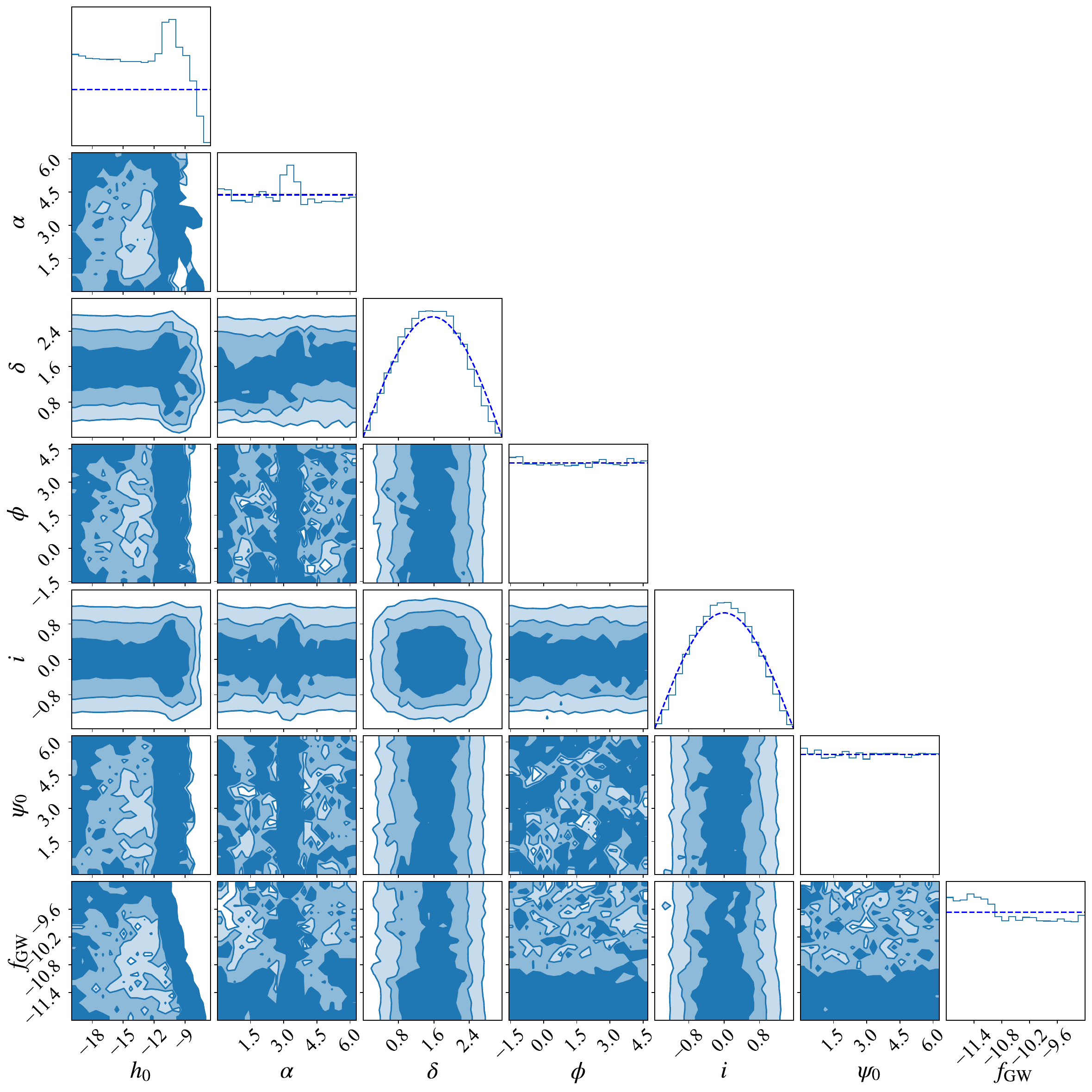}
    \caption{$\dot{P}_b$ search on current data with uniform prior of $f_{\text{GW}}$. Upper limit on $h_0$ can be extracted from the posterior distribution. The priors (\textcolor{blue}{blue dashed line}) are also shown. The contours are set at $50\%, 80\%, 95\%$. All the angles and phase are in radians, and $f_{\text{GW}}$ is in $\log_{10} \text{Hz}$. The following figures follow the same convention.}
    \label{fig:pbdot_corner}
\end{figure*}

\begin{figure*}[ht!]
    \centering
    \includegraphics[width=1\linewidth]{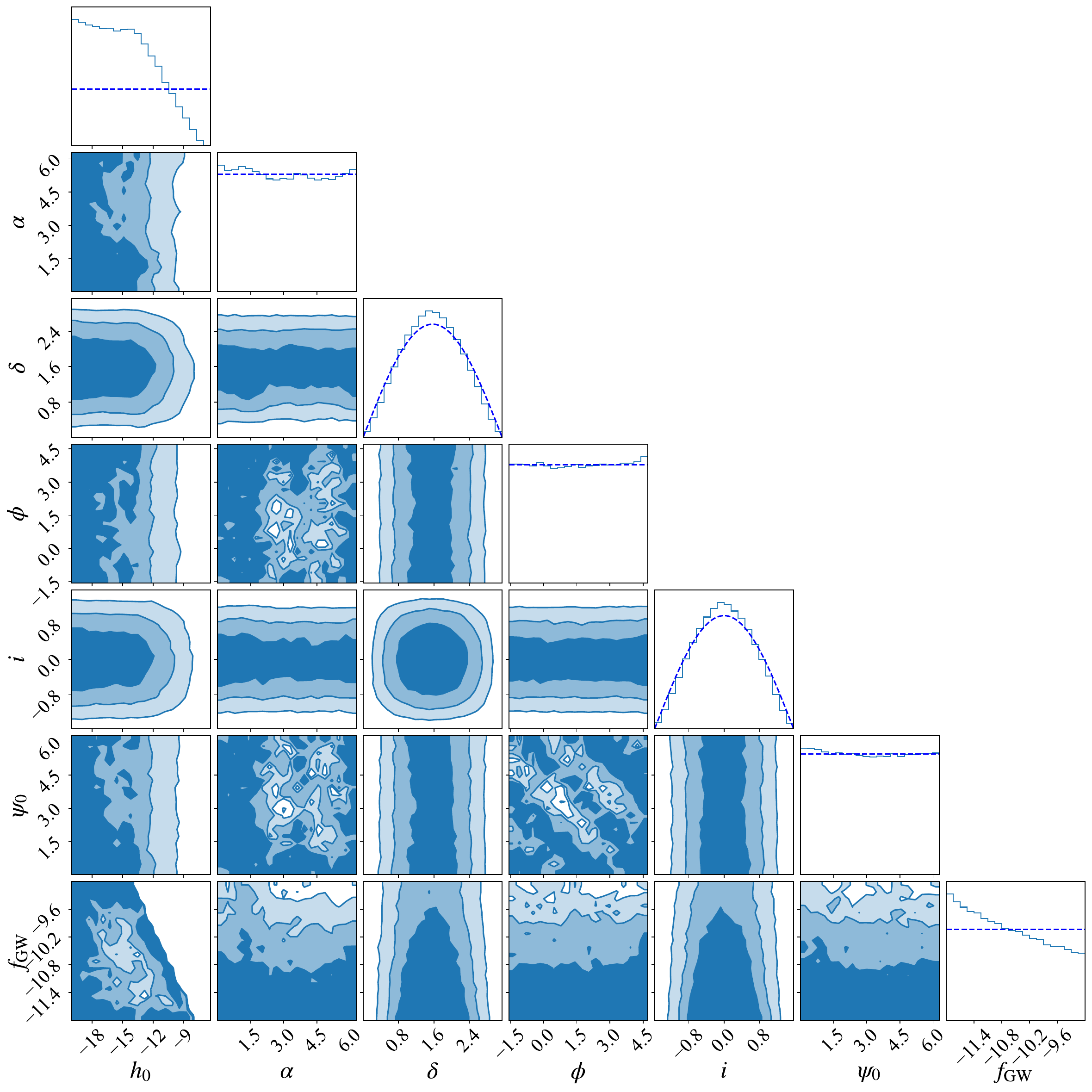}
    \caption{$\ddot{P}$ search on current data with uniform prior of $f_{\text{GW}}$. Upper limit on $h_0$ can be extracted from the posterior distribution. The priors (\textcolor{blue}{blue dashed line}) are also shown.}
    \label{fig:pdd_corner}
\end{figure*}

We then search for CW signals in the real data, running up to $3,000,000$ samples in MCMC and discarding the first $20\%$ as burn-in. For the search we use log uniform prior on $h_0$. No CW signals are found in either data set. Fig.~\ref{fig:pbdot_corner} and Fig.~\ref{fig:pdd_corner} show the posterior distributions of the relevant parameters for $\dot{P}_b$ search and $\ddot{P}$ search respectively. We recover the priors from the analysis except for a constrained strain amplitude. We report $BF=0.6913\pm0.0058$ for the $\dot{P}_b$ search and $BF=0.472 8\pm0.0042$ for the $\ddot{P}$ search. Based on the null distribution and the posterior distributions, we do not detect a signal.

\subsection{Upper limits}
\label{sec:ul}
In the absence of detection, we set upper limits on the CW strain amplitude as a function of frequency by taking the $95\%$ cutoff of the marginalized posterior distribution of $h_0$, as shown in Fig.~\ref{fig:upbd}. For setting the upper limit we adopt a linear uniform prior on $h_0$. Specifically, for each frequency bin we apply a narrow Gaussian prior centered on that frequency with a standard deviation of 0.1 in the log space of frequency. The prior on $h_0$ is chosen to be uniform in linear space, which guarantees that the posterior is agnostic to the lower cut-off of the $h_0$ prior. If a log-uniform prior were used, a smaller lower cut-off in $h_0$ prior would always result in a smaller $h_0$ upper limit.

With our new analytic model, Bayesian framework, and the updated $\dot{P}_b$ data, we improve the $h_0$ upper limits by 9.53 times at a reference frequency of $f_{\text{GW}}=100~\text{pHz}$, an order of magnitude over previous results~\cite{DeRocco2023}. For the current data, the $\ddot{P}$ analysis achieves tighter strain limits than $\dot{P}_b$ for $f_{\text{GW}} \gtrsim 1.5~\text{pHz}$. Nevertheless, a combined analysis across multiple timing parameters remains valuable: when the sensitivity reaches levels comparable to expected CW signals, joint analysis of $\dot{P}_b$, $\ddot{P}$, and potentially other parameters could break degeneracies in source properties including CW frequency and sky locations, providing stronger constraints on strain amplitude. 

\begin{figure*}
\label{fig:skylocation}
\centering
\includegraphics[width=\columnwidth]{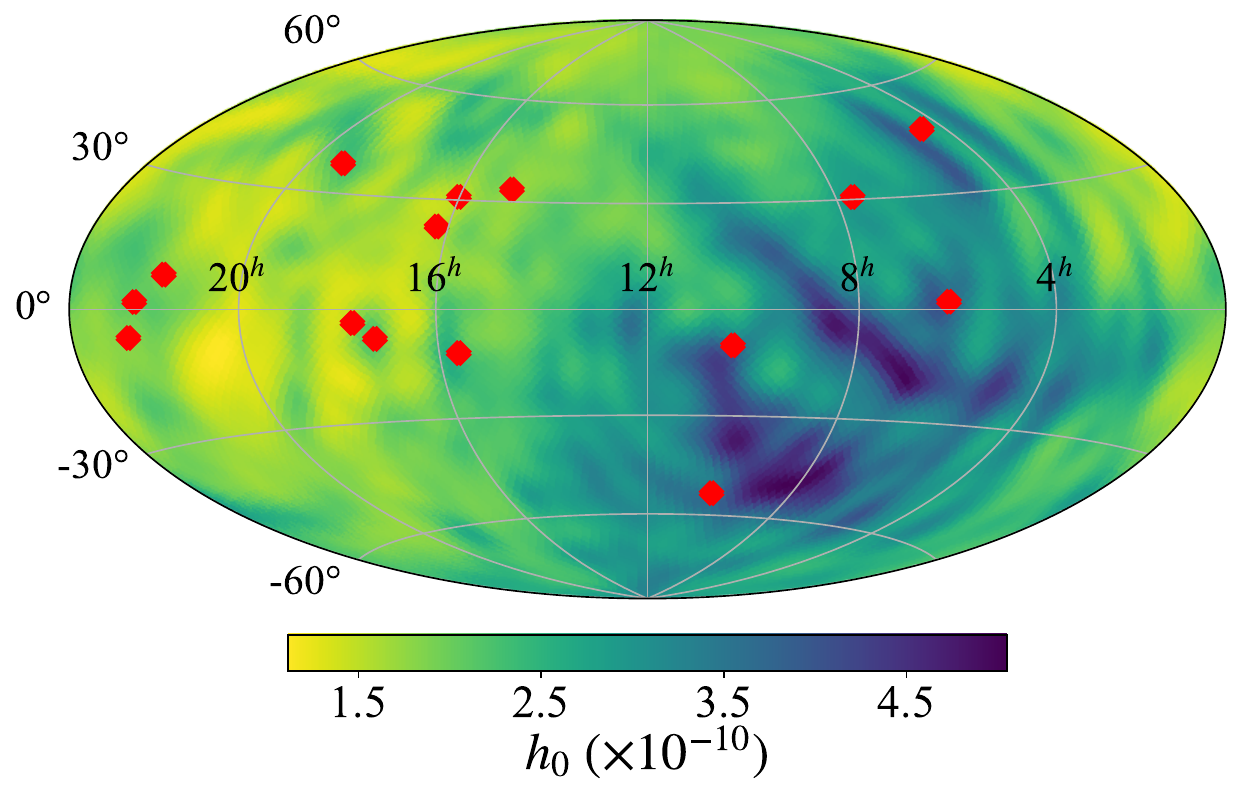}
\includegraphics[width=\columnwidth]{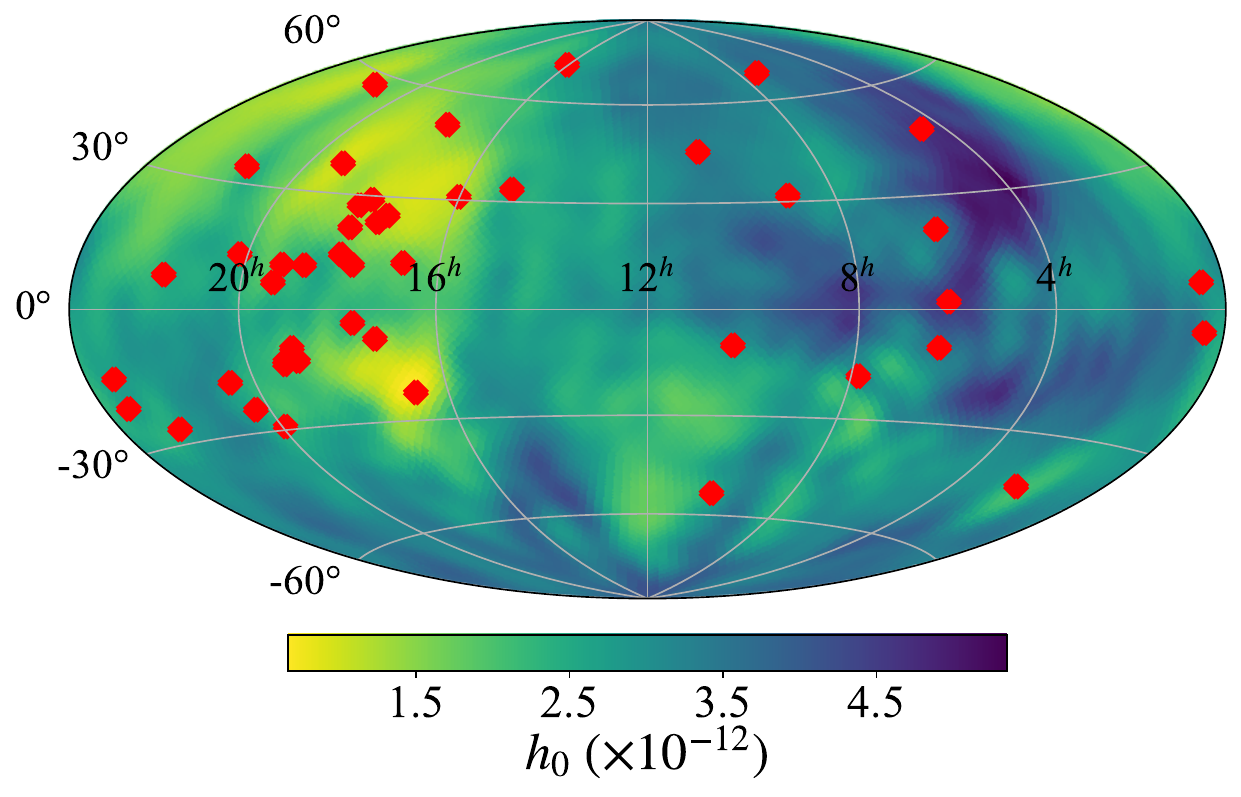}
\caption{$95\%$ upper limit on $h_0$ as a function of sky location, at a reference frequency of $100~\text{pHz}$. Left: $\dot{P}_b$ data. The highest $h_0$ upper limit is $5.1\times10^{-10}$ and the lowest upper limit is $1.1\times10^{-10}$; Right: $\ddot{P}$ data. The highest $h_0$ upper limit is $5.4\times10^{-12}$ and the lowest upper limit is $6.7\times10^{-13}$. The red diamonds denote the locations of the pulsars in the sky. The sensitivity roughly follows the antenna beam pattern.}
\label{fig:skymap}
\end{figure*}

On top of the fully marginalized upper limit, we also calculate the sky location dependent upper limit, presented in Fig.~\ref{fig:skymap}. This is done by fixing $\alpha,\, \delta$ in the Bayesian analysis for the corresponding sky location. In case of no detection, we expect the sensitivity distribution to correlate with the pulsar distribution by virtue of the antenna pattern function defined in Eq.~\ref{eq:antenna}. Our result conforms to this expectation. In the calculation presented here, we use a reference GW frequency of $100~\text{pHz}$. We note that the detection sensitivity can be different by up to an order of magnitude in different regions of sky. The upper limit in strain is as low as $h_0\simeq6.7\times10^{-13}$ for particular sky locations, one seventh of the marginalized result over all sky at the same frequency. 

\subsection{Scaling laws}
The scaling of the upper limit $h_0$ with GW frequency $f_{\text{GW}}$ follows $h_0 \propto f_{\text{GW}}^{-1}$ for $\dot{P}_b$, and $h_0 \propto f_{\text{GW}}^{-2}$ for $\ddot{P}$ (see Fig.~\ref{fig:upbd}). These scaling relations are consistent with our analytic model, where $a_{\text{\text{GW}}} \propto h_0 f_{\text{GW}}$ and $j_{\text{GW}} \propto h_0 f_{\text{GW}}^2$.

At very low frequencies, the scaling of sensitivity changes. For $f_{\text{GW}} \lesssim 10~\text{pHz}$, the GW wavelength exceeds $d_p^{-1}$ for the most sensitive pulsars. In this regime, the earth term and pulsar term in Eq.~\ref{eq:1} become ``in phase," which results in a weakened $h_0$ upper limit for extremely low frequency.

\begin{figure*}[ht!]
    \centering
    \includegraphics[width=0.95\linewidth]{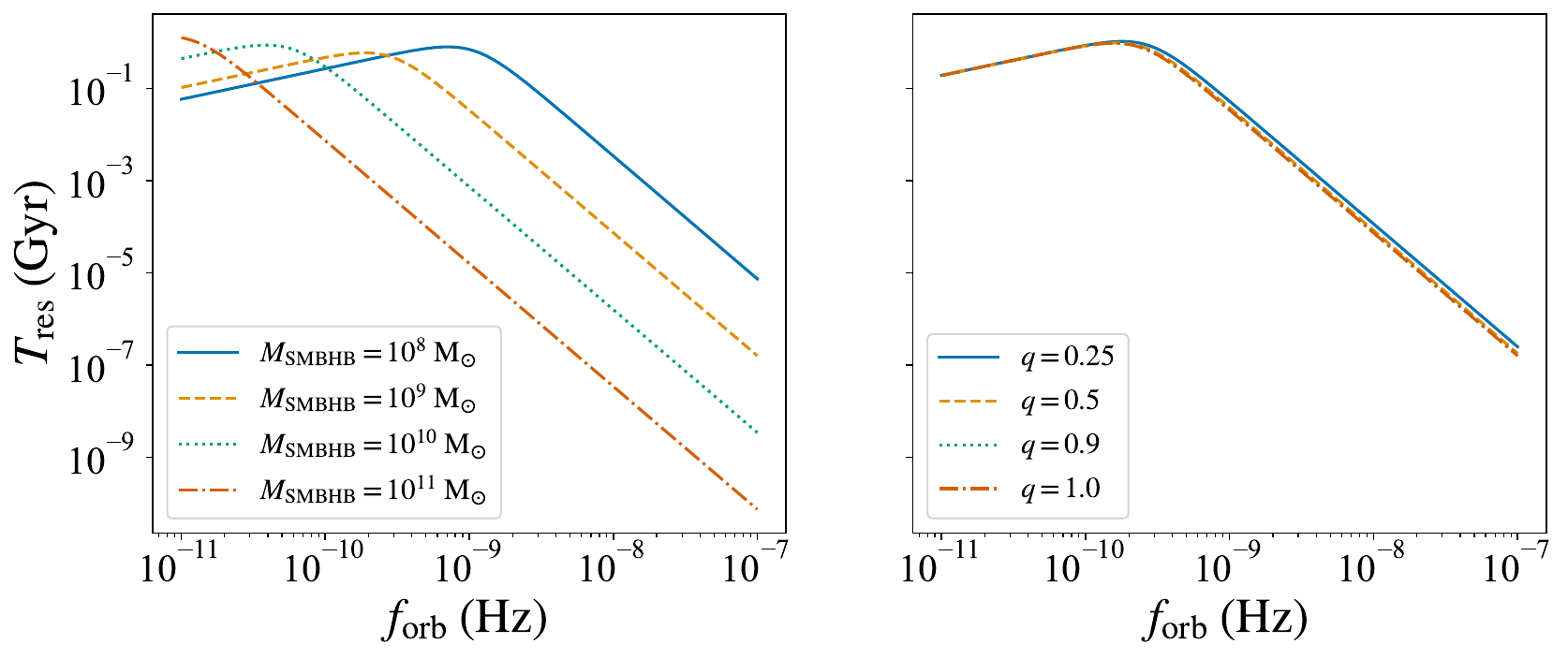}
    \caption{SMBHB residence time $T_{\text{res}}=f_{\text{orb}}/\dot{f}_{\text{orb}}$ as a function of \textbf{orbital} frequencies, for different total masses and mass ratios. The turnover at $\sim \text{nHz}$ corresponds to the turnover in the GWB spectrum in Fig.~\ref{fig:upbd}. LHS: residence time as a function of total SMBHB mass, where mass ratio~$=1$ for all SMBHBs; RHS: residence time as a function of SMBHB mass ratio, where $M_{\text{SMBHB}}=10^9~\mathrm{{M}}_{{\odot}}$. }
    \label{fig:residence_time}
\end{figure*}

\begin{figure}
    \centering
    \includegraphics[width=\columnwidth]{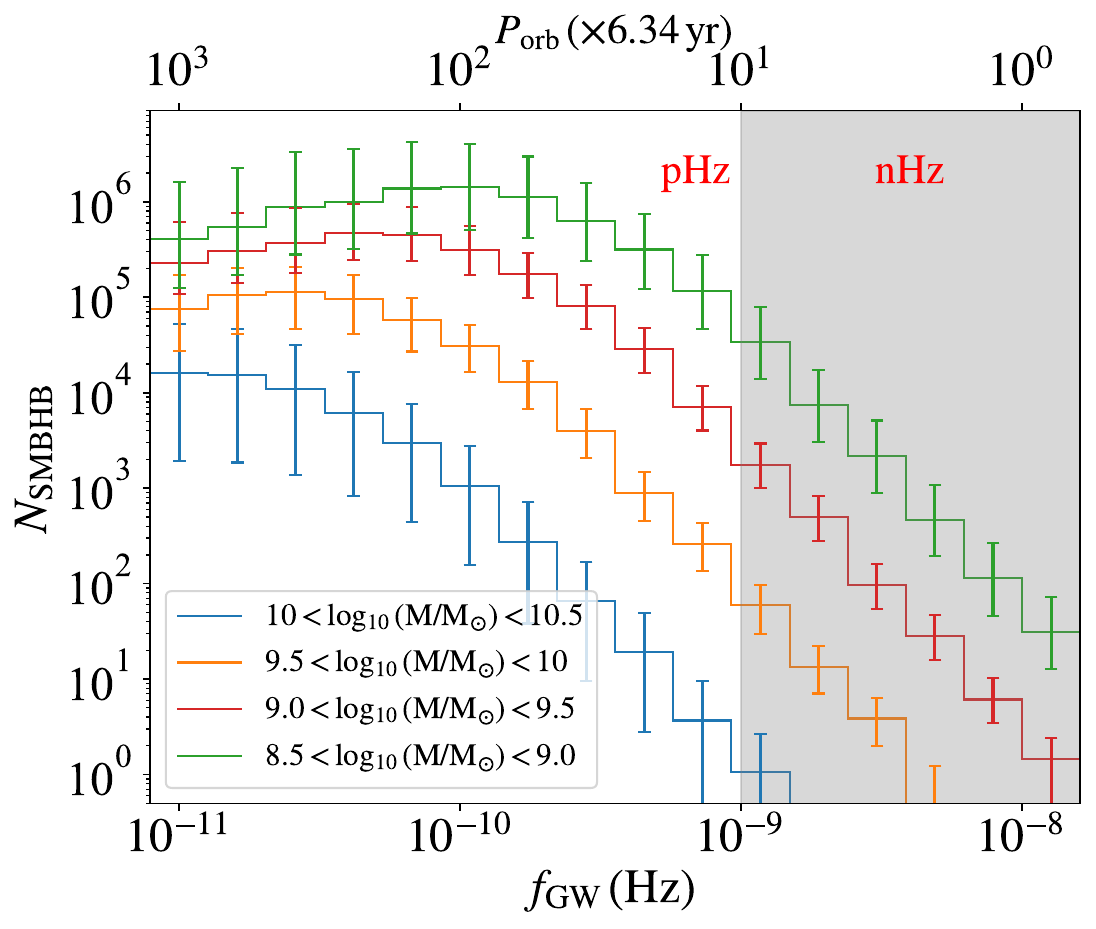}
    \label{fig:SMBHB}
    \caption{The number of SMBHBs per 0.2 log frequency (Hz) as a function of mass range. In pHz regime, SMBHBs across the mass range are likely to contribute to CWs. The SMBHB population is computed by modifying \texttt{oddAGN}~\cite{oddAGN}. There is a turn-over in $N_{\text{SMBHB}}$ at pHz frequencies for all mass ranges considered.}
    \label{fig:N_mass}
\end{figure}

\subsection{SMBHB synthesis model}
\label{sec:SMBHB synthesis}

The integrated GWB spectrum predicted from our SMBHB population model deviates significantly from a simple power-law extrapolation in the sub-nHz regime. In particular, a turnover in the spectrum appears around $60~\text{pHz}$, corresponding to the strongest GWB signal across the entire frequency range. A simple power-law assumes that GW emission dominates SMBHB orbital evolution, while our model incorporates environmental effects such as stellar hardening and dynamical friction. At pHz frequencies, these environmental processes dominate, shortening the binary residence timescale ($f_{\mathrm{orb}}/\dot{f}_{\mathrm{orb}}$) compared to purely GW-driven evolution (Fig.~\ref{fig:residence_time}), where $f_{\mathrm{orb}}$ is the binary orbital frequency. The shorter residence time leads to a suppression of GWB power at the lowest frequencies, manifested as a reduced characteristic strain in Fig.~\ref{fig:upbd}.

We generate $1,000$ realizations of the SMBHB population to estimate the statistical uncertainty, represented by the quantile shading in Fig.~\ref{fig:upbd}. The distribution of SMBHBs across frequency bins for different masses follows a similar pattern (Fig.~\ref{fig:N_mass}), with a turnover occurring near $50~\text{pHz}$, matching the feature in the GWB spectrum. The overall population decreases with increasing SMBHB mass, consistent with the expected mass function.

We extract the loudest individual CW sources in each frequency bin. We find no strong scaling between the strain amplitude of the loudest sources and frequency. To extract these sources, we find the largest strain amplitude from all the bins that contains a SMBHB, for each frequency. The loudest sources typically exhibit strain amplitudes around $1\sim4 \times 10^{-16}$. The majority of individual sources lie several orders of magnitude below this level.

\subsection{Future Detection Prospects}
\label{sec:forecast}
To forecast detection prospects, we conduct simulations to determine the scaling relations between detection sensitivity and (1) data size and (2) data uncertainty levels. To generate the simulated pulsar data, we draw from the extrapolated sky location distribution of known pulsars inspired by method described in~\cite{Xin2021}. Both conservative and more optimistic estimates for SKA1 and 2 are included here. The SKA LOW survey will concentrate on regions within $\pm 5^\circ$ from the galactic plane, while the SKA MID survey will concentrate on regions within $\pm 10^\circ$ of the galactic plane~\cite{Braun:2015zta}. Upcoming experiments will dramatically expand the PTA by increasing the number of precisely timed MSPs to the order of several hundred, representing more than an order-of-magnitude increase in size relative to current data~\cite{Xin2021}. Distance measurements will also improve significantly with SKA astrometry~\cite{Smits:2011}, reducing uncertainties to levels that satisfy our distance-precision criterion for essentially all pulsars. According to~\cite{Xin2021}, SKA1 is expected to produce precise measurements for 675 pulsars, with an additional 120 achieved by SKA2. Because SKA1 consists of southern hemisphere telescopes, we apply a cut-off at $30^\circ$ in the declination angle of the pulsars. We then draw pulsar locations from our empirical sky location distribution extracted from all the pulsars in IPTA data~\cite{Antoniadis2022}.

\begin{figure*}[ht!]
    \centering
    \includegraphics[width=\columnwidth]{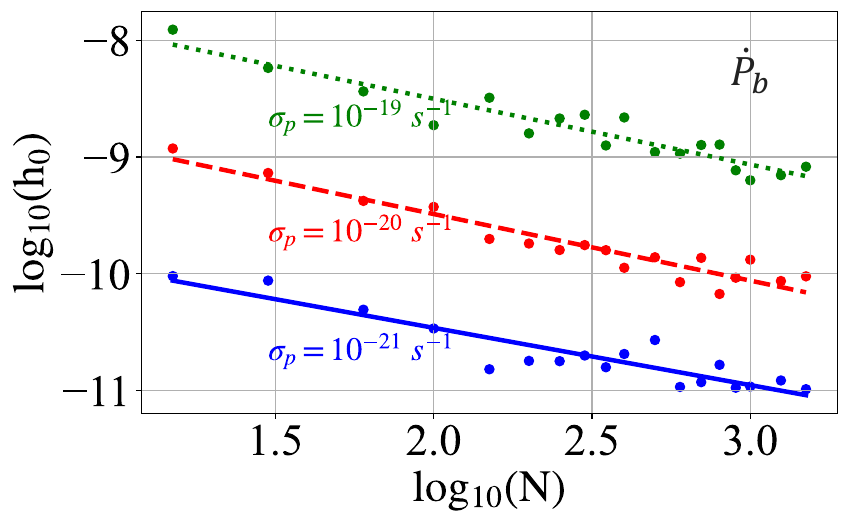}
    \includegraphics[width=\columnwidth]{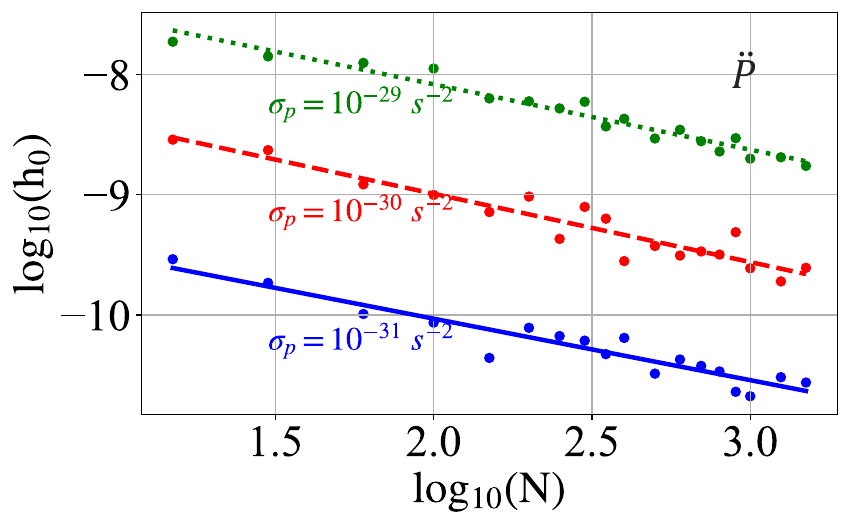}
    \caption{Scaling of detection sensitivity measured by detectable CW strain amplitude as a function of data size N, calculated at a reference frequency of $10~\text{pHz}$ as a conservative estimate, for both $\dot{P}_b$ and $\ddot{P}$ analyses. For higher frequencies, the sensitivity improves. For $\dot{P}_B$ analysis, $\sigma_p$ denotes the uncertainty in $\Delta\dot{P}_b/P_b$, which is the observed quantity that should be equal to the CW dependent $a_{\text{GW}}$; for $\ddot{P}$ analysis, $\sigma_p$ denotes the uncertainty in $\ddot{P}/P$, which is the observed quantity that should be equal to the CW dependent $j_{\text{GW}}$. The fitted lines have slopes of roughly -0.5, which agrees with the expected scaling law with N. For any larger frequencies, the method is sensitive to lower strain amplitude: for $\dot{P}_b$ analysis, $h_0\propto f_{\text{GW}}^{-1}$; for $\ddot{P}$ analysis, $h_0\propto f_{\text{GW}}^{-2}$}.
    \label{fig:scaling}
\end{figure*}

We verify that, when probing strain amplitudes above potential CW sources, uncertainties in pulsar distances alone do not significantly impact the upper limits on $h_0$. Each simulation assumes a null CW signal. We find the detection sensitivity scales as $\propto \sigma_p/\sqrt{N}$ when $N$ is asymptotically large (above $\sim100$), where $\sigma_p$ denotes the uncertainty in data used for inference and $N$ is the number of pulsars (Fig.~\ref{fig:scaling}). When $N$ is below 50, boundary effects kick in and the scaling deviates from $\propto 1/\sqrt{N}$---the exponent on $N$ in the power law scaling is slightly smaller than $-1/2$.

\begin{figure}
\centering
\includegraphics[width=0.95\columnwidth]{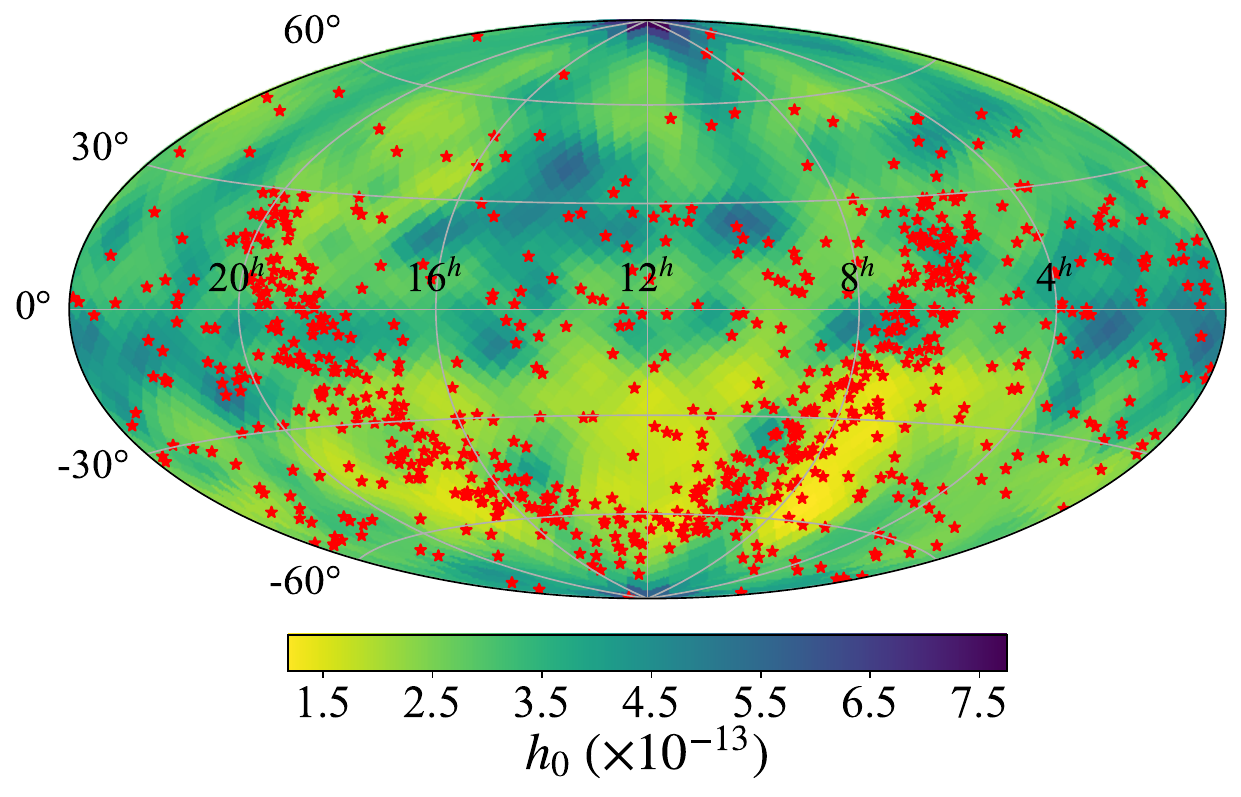}
\caption{Forecast sky-location dependent strain upper limits in SKA era at a reference frequency $f_{\text{GW}}=100~\text{pHz}$. On top of the pulsars already timed by EPTA, NANOGrav and PPTA, the sky locations of the SKA pulsars are drawn from a distribution extrapolated based on currently known pulsars. The highest $h_0$ upper limit is $7.8\times10^{-13}$ and the lowest upper limit is $1.3\times10^{-13}$.}
\label{fig:forecast_sky}
\end{figure}

For a conservative forecast, we assume each pulsar retains the typical uncertainty level seen in the current data. We further assume that $80\%$ of the $\sim800$ precisely timed pulsars are in binary systems~\cite{Lorimer2008}, each with a measurable $\dot{P}_b$, and that all have measurable $\ddot{P}$. In addition, we take the observation time for all the new pulsars to be 10 years, which is below the longest observation time in the current data. Under these assumptions, we compare the forecast sensitivities derived from $\dot{P}_b$ and $\ddot{P}$ analyses with the strain amplitudes of the loudest individual sources from our SMBHB population model. As shown in Fig.~\ref{fig:upbd}, SKA-based data will enable an additional order-of-magnitude improvement over current sensitivities, solely due to the increase in data size. Our simulations also show that with more pulsars, the detection sensitivity will become more isotropic (Fig.~\ref{fig:forecast_sky}). SKA data is expected to reduce the variance between the most and least sensitive sky locations by $\sim40\%$ for $\ddot{P}$ analysis. However, the strain amplitudes of individual SMBHB CW sources drawn from our GWB synthesis model are still $\sim3$ orders of magnitude below this forecast sensitivity.

\subsection{Injection of a CW}
\label{sec:individual_source}

To illustrate that our method is not only able to set upper limits, but capable of detecting an individual CW source, we draw the loudest SMBHB from our population and simulate its GW imprints on the timing parameters. We then retrieve the CW signal from the simulated data using our method. We focus on $\ddot{P}$ in this section and predict the required precision of $\ddot{P}$ measurement for a realistic detection.

\begin{figure}[ht!]
    \centering
    \includegraphics[width=0.92\columnwidth]{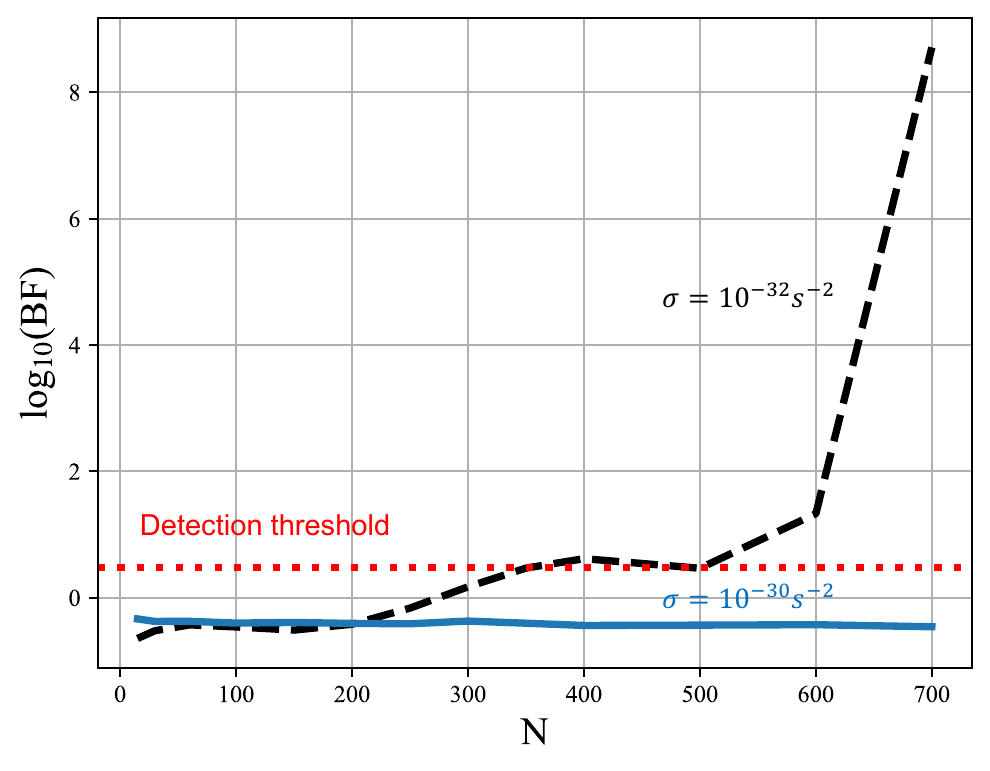}
    \caption{The scaling between Bayes factors favoring the existence of the injected CW signal over null and number of pulsars N in the simulated data. The results being plotted are for $\ddot{P}$ analysis, which is more promising for a detection. The detection threshold corresponds to the $95\%$ cut-off in the null BF distribution. $\sigma_p$ is the typical uncertainty in the measured $\ddot{P}/P$ for a pulsar in the dataset.}
    \label{fig:N_scaling}
\end{figure}

\begin{figure}[b]
\centering
\includegraphics[width=0.92\columnwidth]{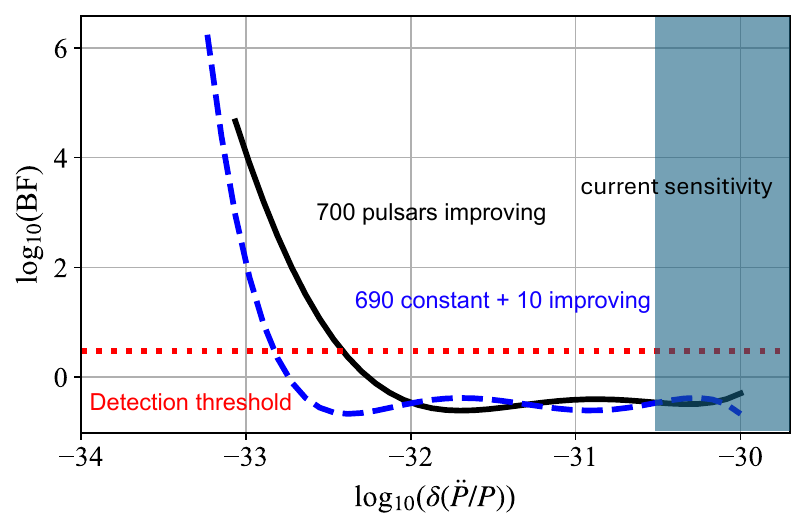}
\caption{Two strategies to achieve a detection: (a) Improve the precision of $\ddot{P}$ measurement for all the pulsars; (b) Improve the precision of $\ddot{P}$ measurement for a selected few pulsars. Simulations show that an improvement of 2.5 (3) orders of magnitude is necessary for a detection in the former (latter) case in data uncertainty. The detection threshold is set by the $95\%$ cut-off in the BF null distribution.}
\label{fig:sigma_scaling}
\end{figure}

In our simulations, we choose NGC 6240 as the host galaxy of the SMBHB. NGC 6240 is a nearby ultraluminous infrared galaxy and prominent candidate for dual AGN, which in turn strongly indicates the presence of a SMBHB~\cite{sobolenko2022ngc6240}. We assume the CW from the SMBHB has the strongest strain amplitude predicted by our population model and a redshifted frequency of $300~\text{pHz}$.

We use BF as an indicator of detection. In Fig.~\ref{fig:N_scaling}, we show how the BF scales with the number of pulsars. The blue line assumes the current uncertainty level in the data, the black dashed line assumes an universal improvement in the uncertainty of measured $\ddot{P}$ by 2 orders of magnitude in all pulsars, and the red dotted line is the detection threshold set by the null distribution. It is evident that the increasing number of pulsars alone by SKA will not facilitate a detection---more precise measurements of pulsar timing parameters are needed.

To forecast detection prospects with improving data uncertainty, we carry out simulations under two scenarios, as shown in Fig.~\ref{fig:sigma_scaling}. The first is when the parameter measurements for all the pulsars improve at the same rate, plotted as a black solid line, while the second is considering a scenario where 10 pulsars have improving data uncertainty and the remaining pulsars have their precision fixed at the typical uncertainty in current data. Our results suggest a uniform improvement of two orders of magnitude on the existing sensitivity ($\delta(\ddot{P}/P)\sim\times10^{-30} s^{-2}$) is necessary, or, alternatively a three order of magnitude improvement for 10 out of the 795 pulsars. So there is a clear trade-off between precision required and number of pulsars that can achieve this precision---the larger the number of pulsars achieving the precision, the lower the precision requirement. Though it may seem as if a uniform improvement of two orders of magnitude would be easier to achieve, it may actually prove the case that an improvement on only 10 pulsars is more realistic. Measuring $\ddot{P}$ requires extremely stable pulsars and long timing baselines, hence it is likely that even with SKA data, only a select few pulsars will be able to provide precise measurements of their $\ddot{P}$.   

In Appendix~\ref{app:cw detection}, we show the detailed search results of the two cases discussed above, along with the results of a search using 795 pulsars with typical uncertainty of current $\ddot{P}$ data, in which we detect no signal (Fig. \ref{fig:700_current}). Fig.~\ref{fig:10_improved} assumes that 10 of these pulsars achieve 3 orders of magnitude improvement in $\ddot{P}$ uncertainty, and is able to recover $h_0$ fairly well. This search also produces posteriors with peaks near the true values for other parameters, such as $f_{\mathrm{GW}}$ and SMBHB sky location. The high degrees of degeneracies between the various parameters lead to multiple peaks in the posteriors. We verify that even smaller uncertainties in the data eventually make the posteriors converge to the injected parameter values. Fig.~\ref{fig:700_improved} assumes that all of 795 pulsars achieve 3 orders of magnitude improvement in $\ddot{P}$ uncertainty, and reaches better convergence to the true values of the injected CW parameters.

In summary, we find that to detect a realistic CW source, a $\sim2$ order of magnitude improvement in the uncertainty of $\ddot{P}$ is required for all pulsars that SKA will observe. Interestingly, however, a trade-off exists between the number of well-measured $\ddot{P}$s and the required precision; in particular, we show only 10 out of 795 pulsars in SKA would need to improve by $\sim 3$ orders of magnitude in order to achieve a detection. In which direction this trade-off will play out when SKA begins taking data remains an open question.

\section{Discussion and Conclusions}
\label{sec:discuss}

As difficult as the required improvement in timing parameter measurement for a future detection may seem, multiple factors work in our favor. Current studies indicate the timing precision of SKA could be up to 2 orders of magnitude better than today~\cite{liu2014pulsar}. On top of that, it has been empirically shown that a longer observation time reduces the standard deviations in timing parameters~\cite{andrews2020nanograv}, and simulations suggest a scaling of $\sim T_{\mathrm{obs}}^{-1/2}$~\cite{liu2019correlated}. The typical timing precision of current operational PTAs is $\sim 1~\mu s$, and the current uncertainty of the period normalized $\dot{P}_b/P_b$, $\ddot{P}/P$ sits at $10^{-20} s^{-1}$, $10^{-30} s^{-2}$. Optimistically the uncertainties in the timing parameters scale linearly with the timing precision. With 2 orders or magnitude improvement in the timing precision, we would achieve $\sigma(\dot{P}_b/P_b)\sim 10^{-22}s^{-1}$\, /\, $\sigma(\ddot{P}/P)\sim 10^{-32}s^{-2}$. The observation times of current PTAs range from a few years to two decades. Longer timespan will only improve the uncertainty in timing parameters, though the improvement due to longer timespan will be incremental compared to the improvement due to better timing precision by the scaling relations. Referring to Fig~\ref{fig:sigma_scaling}, it is likely that some potential CW sources may fall within our sensitivity range with SKA data. 

However, there are a few caveats. Currently, the dominant sources of uncertainty in both $\dot{P}_b$ and $\ddot{P}$ are due to the measured PTA timing parameters, which are expected to improve with longer time baselines and next-generation telescopes such as SKA. As timing precision improves, however, red noise and uncertainties in galactic acceleration will not be negligible. These effects present their own challenges and motivate the exploration of higher-order timing parameters with better-controlled uncertainties. More details of recent developments can be found for red noise mitigation in~\cite{Goncharov_2020} and for Galactic acceleration modeling in~\cite{ibata2023chartinggalacticaccelerationfield}. In particular, since red noise is covariant with a GWB signal, it will be interesting to include a GWB when extracting timing parameters, which has not been done. We also acknowledge that only a very small fraction of pulsars will have precise measurements of higher-order timing parameters, due to the stringent requirement of pulsar stability and long timing baseline. However, we have shown in Sec.~\ref{sec:individual_source} that a small number of precise measurements could potentially facilitate a successful detection.
 
Targeted searches may also help improve detection prospects~\cite{Agarwal:2025cag}. In our framework, targeted searches are achieved by adopting informative priors on CW parameters. These priors are often informed by multi-messenger astrophysical observations. For example, recent work has shown that dual AGNs can serve as tracers of SMBHBs, where the sky location of candidate dual AGNs can be determined through optical, radio, and X-ray observations, while redshifts are measured spectroscopically~\cite{Casey-Clyde2022}. Moreover, the orientation of SMBHB candidates can be constrained through observations of jet alignment~\cite{Dotti2010}, broad-line region kinematics~\cite{Afanasiev2019}, variability and polarization of quasar light~\cite{Hutsemekers2005}, and host galaxy morphology~\cite{Zakamska2006}.

The framework presented here can also be used to search for alternative deterministic GW signals by adapting the CW model parameters. This opens opportunities to search for a variety of exotic physics, including ultralight dark matter and dark matter substructure~\cite{Brito2015, NANOGrav2023NewPhysics}. 

The exact effect of gas hardening on binary formation and evolution is still an open research question. For our GWB synthesis model, the energy loss of the binary by gas hardening is likely to be comparable with GW emission and stellar hardening at a transitional binary separation of $\sim1$ pc~\cite{Koss2023UGC4211, Izquierdo‐Villalba2021}. Below this separation, the energy loss is dominated by GW emission; above this separation, energy loss is dominated by stellar hardening. Inclusion of gas hardening in the model may increase the turnover frequency in the GWB spectrum, since it will further reduce the residence time of a binary and thus GWs emitted at a certain frequency. Towards the very beginning of binary formation, gas hardening is expected to be subdominant to stellar hardening and dynamical friction, so the low-frequency asymptotic behavior of our GWB spectrum should remain unaltered. 

Looking forward, there are several promising extensions to pursue. First, a dedicated analysis targeting the pHz GWB could provide independent constraints on the properties of sources such as SMBHBs, topological defects, or early-Universe phase transitions. Second, a key technical challenge in CW searches is the presence of a background signal, which can act as a source of contamination. Future work should incorporate the treatment of an unresolved GWB as a correlated noise component when searching for individual CW sources, a step necessary for maintaining sensitivity in a background-dominated regime. Third, the joint analysis of multiple timing parameters, such as $\dot{P}_b$, $\ddot{P}$, and higher-order derivatives, offers a pathway to break degeneracies between CW source properties and enhance detection significance. Finally, extending the framework to accommodate alternative GW morphologies, including non-monochromatic signals and bursts with memory, would substantially broaden the range of astrophysical and cosmological phenomena accessible to pHz GW searches. Each of these future developments is essential for maximizing the scientific return of next-generation PTA data and for exploiting the full potential of this method. 

In summary, we have carried out the most comprehensive search to date for pHz gravitational waves using secular pulsar parameters, achieving an order-of-magnitude improvement in strain limits over earlier studies. By introducing a Bayesian detection framework, doubling the pulsar sample, and connecting the analysis directly to population synthesis models, we establish this approach as a robust tool for probing the ultra-low frequency regime. Looking ahead, the dramatic increase in pulsar discoveries and timing precision with SKA will transform these methods from upper-limit exercises into genuine discovery opportunities. This opens a realistic path to testing models of supermassive black hole binary evolution, while simultaneously constraining exotic physics such as topological defects and early-Universe phase transitions.

\acknowledgements 
We thank Jeffrey S. Hazboun for the discussion of SKA pulsar population forecasting. The research of J.D. is supported in part by the U.S. Department of Energy grant number DE-SC0025569. W.D. was supported by NSF grant PHY-2210361 and the Maryland Center for Fundamental Physics.

\clearpage

\appendix
\section{Derivation of analytic model}
\label{sec:analytic derivation}

GWs are ripples of spacetime, and thus will induce an apparent velocity between the pulsars and the solar system barycenter. This, and its derivatives, is the main quantity used to detect a signal of GW in this method discussed. Here we derive the observable shifts in timing parameters due to the CWs emitted by a SMBHB, which is assumed to be the main source of CW in this paper. The general idea can be adapted to model other sources.

We start from the well known expression

\begin{dmath}
    v_{GW}=\sum_{A=+,\times} F_A(\hat{\Omega})(h_A(t,0)-h_A(t-d_p,\vec{d_p})=\frac{\hat{d_p^i}\hat{d_p^j}}{2(1+\hat{\Omega}\cdot \hat{d_p})}[e_{ij}^{+}(\hat{\Omega})(h_{+}(t,0)-h_{+}(t-d_p,\vec{d_p}))+e_{ij}^{\times}(\hat{\Omega})(h_\times(t,0)-h_{\times}(t-d_p,\vec{d_p}))] \, ,
\end{dmath}

where $\hat{\Omega}$ is the unit vector of GW propagation direction, while $\hat{d_p}$ is the vector pointing to the pulsar under consideration. We assume the CWs are emitted by the fiducial circular orbit SMBHBs

\begin{equation}
    h_0=\frac{2(G\mathscr{M})^{5/3}}{d_L}\omega_0^{2/3}
\end{equation}

where $\mathscr{M}$ is the chirp mass, $d_L$ is the luminosity distance. 

\begin{dmath*}
    \begin{cases}
        h_+=h_0\cos(2\phi)(1+\cos^2(i))\sin(2\Psi)\\~~~~~~~~+2h_0\sin(2\phi)\cos(i)\cos(2\Psi)\\
        h_\times=h_0\sin(2\phi)(1+\cos^2(i))\sin(2\Psi)\\~~~~~~~~-2h_0\cos(2\phi)\cos(i)\cos(2\Psi)\\
    \end{cases}
\end{dmath*}

where $\phi$, $i$ defines the orientation of the GW source, and $\Psi=\Psi_0+\omega_0t$, with $\omega_0$ the orbital frequency of the binary gravitational wave source.

Expand by virtue of equation 2, we have

\begin{widetext}
\begin{dmath*}
    v_{GW}=\frac{\hat{d_p^i}\hat{d_p^j}}{2(1+\hat{\Omega}\cdot \hat{d_p})}[e_{ij}^{+}(\hat{\Omega})(h_0\cos(2\phi)(1+\cos^2(i))\sin(2\Psi_0+2\omega_0t)+2h_0\sin(2\phi)\cos(i)\cos(2\Psi_0+2\omega_0t)-h_0\cos(2\phi)(1+\cos^2(i))\sin(2\Psi_0+2\omega_0(t-\hat{\Omega}\cdot\vec{d_p}))-2h_0\sin(2\phi)\cos(i)\cos(2\Psi_0+2\omega_0(t-\hat{\Omega}\cdot\vec{d_p})))+e_{ij}^{\times}(\hat{\Omega})(h_0\sin(2\phi)(1+\cos^2(i))\sin(2\Psi_0+2\omega_0t)-2h_0\cos(2\phi)\cos(i)\cos(2\Psi_0+2\omega_0t)-h_0\sin(2\phi)(1+\cos^2(i))\sin(2\Psi_0+2\omega_0(t-\hat{\Omega}\cdot\vec{d_p}))+2h_0\cos(2\phi)\cos(i)\cos(2\Psi_0+2\omega_0(t-\hat{\Omega}\cdot\vec{d_p})))]
\end{dmath*}
\end{widetext}

The direction of the source can be parametrized by the right ascension $\alpha$ and declination $\delta$. In the Cartesian coordinates,

\begin{equation}
    \begin{cases}
        \hat{\alpha}=-\sin(\alpha) \hat{i}+\cos(\alpha) \hat{j}\\
        \hat{\delta}=\cos(\delta) \cos(\alpha) \hat{i} + \cos(\delta) \sin(\alpha) \hat{j}-\sin(\delta) \hat{k}
    \end{cases}
\end{equation}

and

\begin{equation}
    \begin{cases}
        e_+=\hat{\alpha}\hat{\alpha}-\hat{\delta}\hat{\delta}\\
        e_\times=\hat{\alpha}\hat{\delta}+\hat{\delta}\hat{\alpha}
    \end{cases}
\end{equation}

In the analysis, we need to use the acceleration and jerk $a_{GW}$, $j_{GW}$, which we derive here. Since we are working in the pHz frequency range with typical periods much longer than the observation time, we can safely assume that $\omega_0 t << 1$ in our derivations, and thus expand $\cos(2\omega_0t)$ as 1, and $\sin(2\omega_0t)$ as $2\omega_0t$. Now we have

\begin{widetext}
\begin{dmath}
    a_{GW}=\frac{\hat{d_p^i}\hat{d_p^j}}{1+\hat{\Omega}\cdot \hat{d_p}}\omega_0 h_0[e_{ij}^{+}(\hat{\Omega})[\cos(2\phi)(1+\cos^2(i))(\cos(2\Psi_0)-\cos(2\Psi_0-2\omega_0(1+\hat{\Omega}\cdot \hat{d_p})d_p))+2\sin(2\phi)\cos(i)(-\sin(2\Psi_0)+\sin(2\Psi_0-2\omega_0(1+\hat{\Omega} \cdot \hat{d_p})d_p))]+e_{ij}^{\times}(\hat{\Omega})[\sin(2\phi)(1+\cos^2(i))(\cos(2\Psi_0)-\cos(2\Psi_0-2\omega_0(1+\hat{\Omega}\cdot \hat{d_p})d_p))-2\cos(2\phi)\cos(i)(-\sin(2\Psi_0)+\sin(2\Psi_0-2\omega_0(1+\hat{\Omega}\cdot \hat{d_p})d_p))]]
\end{dmath}
\end{widetext}

\begin{widetext}
\begin{dmath}
    j_{GW}=-2\frac{\hat{d_p^i}\hat{d_p^j}}{1+\hat{\Omega}\cdot \hat{d_p}}\omega_0^2 h_0[e_{ij}^{+}(\hat{\Omega})[\cos(2\phi)(1+\cos^2(i))(\sin(2\Psi_0)-\sin(2\Psi_0-2\omega_0(1+\hat{\Omega}\cdot \hat{d_p})d_p))+2\sin(2\phi)\cos(i)(\cos(2\Psi_0)-\cos(2\Psi_0-2\omega_0(1+\hat{\Omega} \cdot \hat{d_p})d_p))]+e_{ij}^{\times}(\hat{\Omega})[\sin(2\phi)(1+\cos^2(i))(\sin(2\Psi_0)-\sin(2\Psi_0-2\omega_0(1+\hat{\Omega}\cdot \hat{d_p})d_p))-2\cos(2\phi)\cos(i)(\cos(2\Psi_0)-\cos(2\Psi_0-2\omega_0(1+\hat{\Omega}\cdot \hat{d_p})d_p))]]
\end{dmath}
\end{widetext}

The unit vectors $\hat{d_p}$ and $\hat{\Omega}$ can be parametrized by the equatorial ascension($\alpha_p, \alpha$) and declination ($\delta_p, \delta$) of the pulsars and source in the Cartesian coordinates, which read

\begin{equation}
    \begin{cases}
        \hat{d_p}=\sin(\delta_p)\cos(\alpha_p)\hat{i}+\sin(\delta_p)\sin(\alpha_p)\hat{j}+\cos(\delta_p)\hat{k}\\
        \hat{\Omega}=-\sin(\delta)\cos(\alpha)\hat{i}+\sin(\delta)\sin(\alpha)\hat{j}+\cos(\delta)\hat{k}
    \end{cases}
\end{equation}

One can then write down explicit expressions for $a_{\text{GW}}, j_{\text{GW}}$ in terms of the parameters above, for which the numerical values are given in the data.

\section{CW detection}
Detailed search results using simulated data with CW signal injected, as described in Sec.~\ref{sec:individual_source}.
\label{app:cw detection}

\begin{figure*}[h!]
    \centering
    \includegraphics[width=\linewidth]{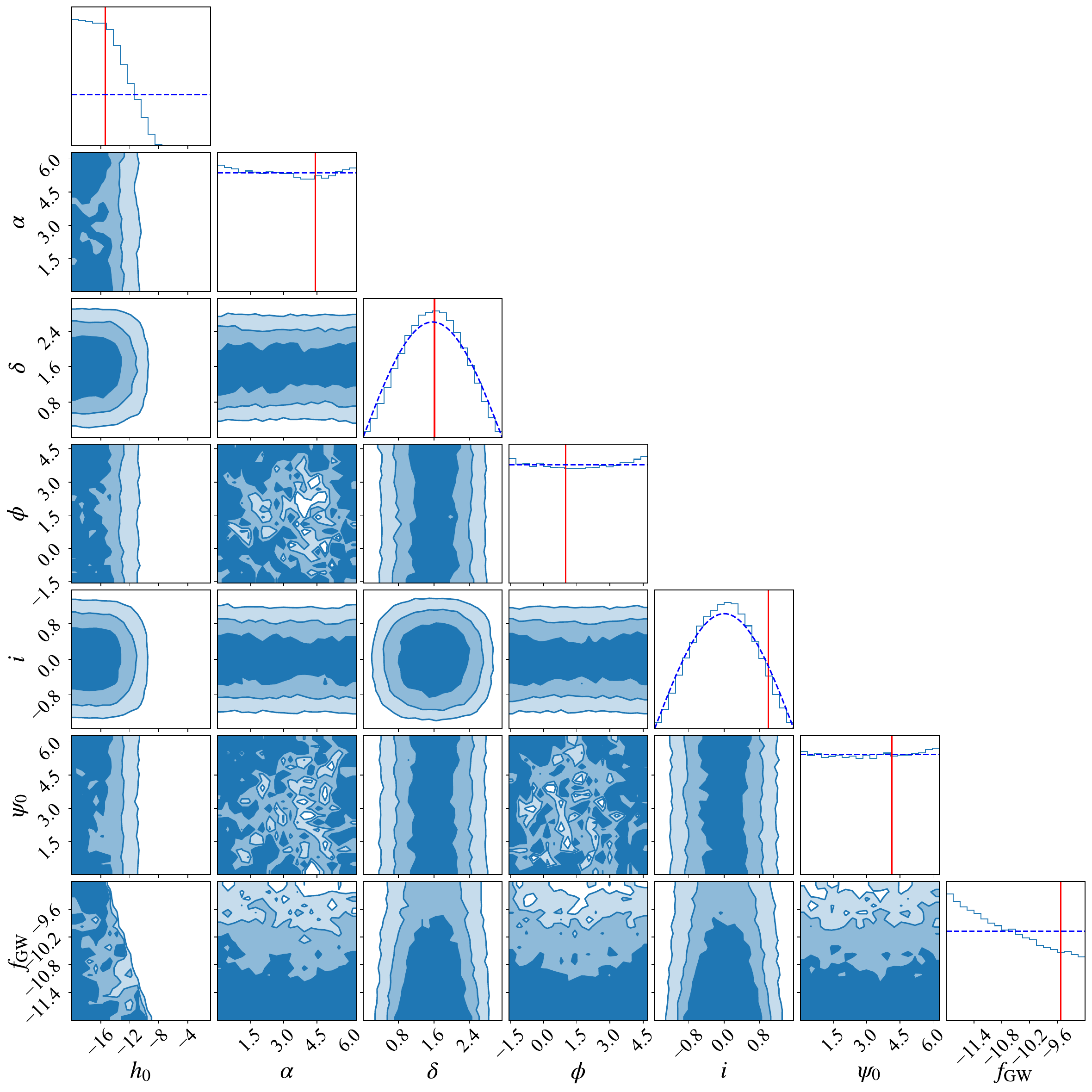}
    \caption{Search for CW source described in Sec.~\ref{sec:individual_source} with uniform priors, 795 pulsars expected from SKA, and current $\ddot{P}$ measurement uncertainty. The injected true values (\textcolor{red}{red solid line}) and priors (\textcolor{blue}{blue dashed line}) are also shown.}
    \label{fig:700_current}
\end{figure*}

\begin{figure*}[h!]
    \centering
    \includegraphics[width=1\linewidth]{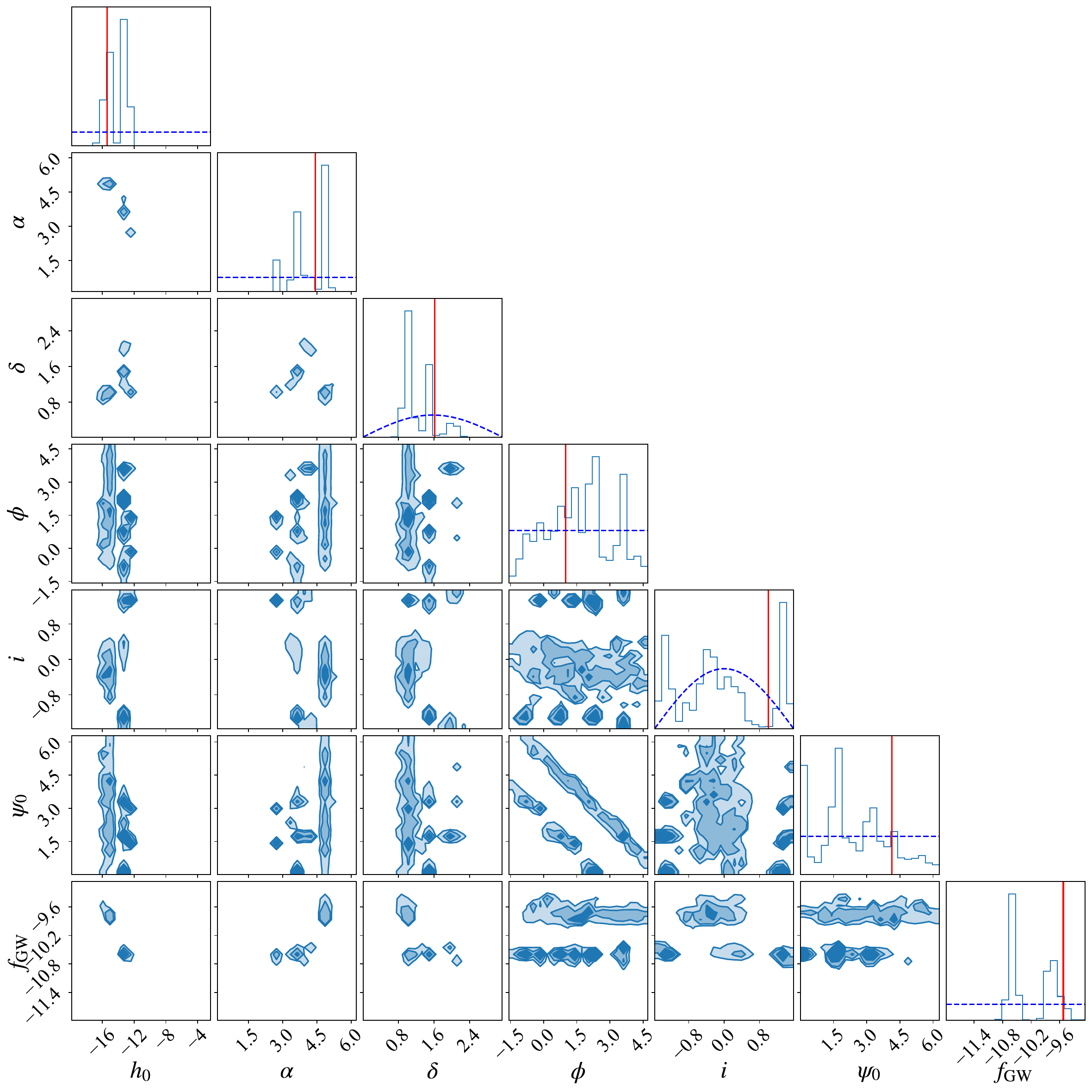}
    \caption{Search for CW source described in Sec.~\ref{sec:individual_source} with uniform priors, 795 pulsars expected from SKA, and current $\ddot{P}$ measurement uncertainty except for 10 pulsars with 3 orders of magnitude improvement in the uncertainty. The injected true values (\textcolor{red}{red solid line}) and priors (\textcolor{blue}{blue dashed line}) are also shown.}
    \label{fig:10_improved}
\end{figure*}

\begin{figure*}[h!]
    \centering
    \includegraphics[width=1\linewidth]{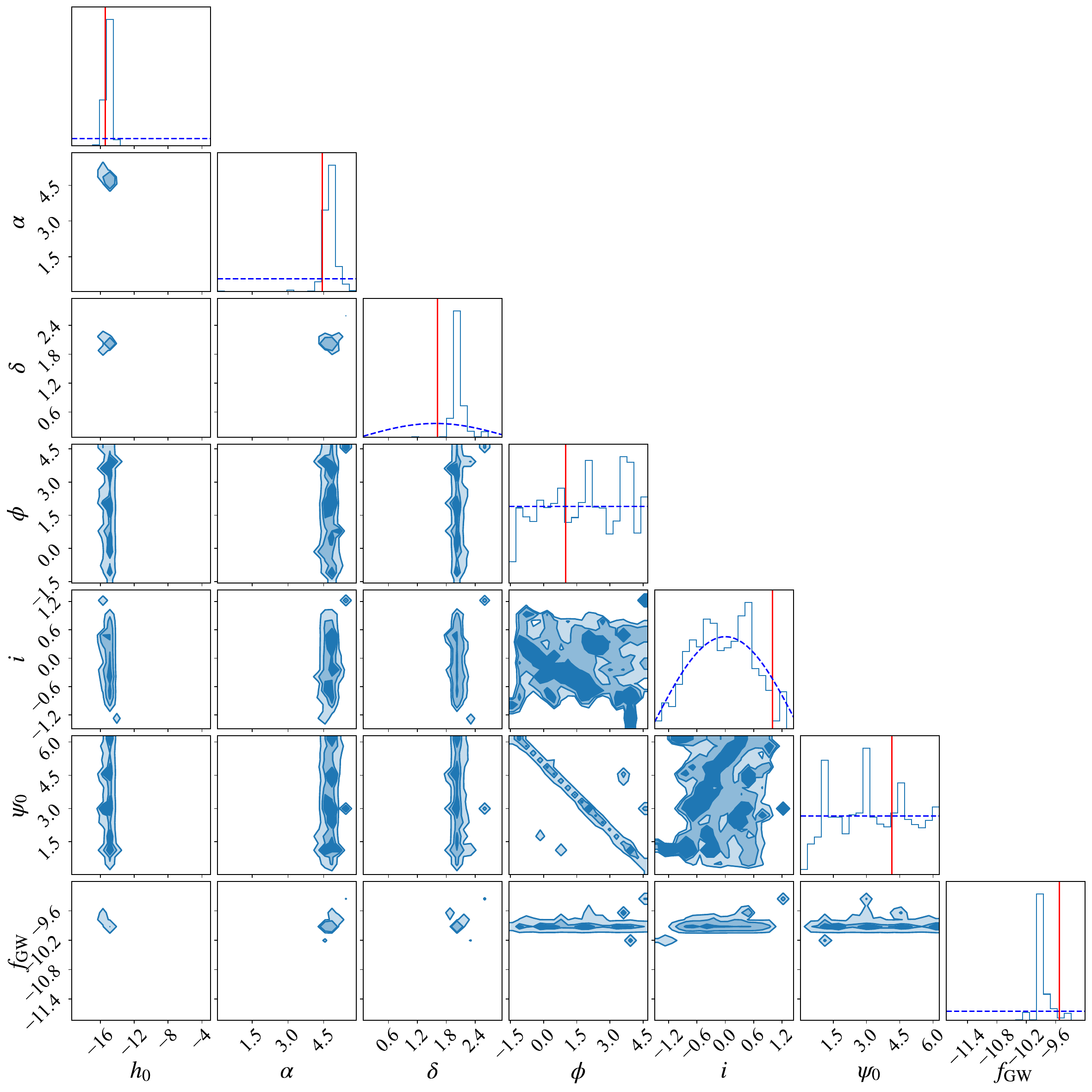}
    \caption{Search for CW source described in Sec.~\ref{sec:individual_source} with uniform priors, 795 pulsars expected from SKA, and 3 orders of magnitude improvement in $\ddot{P}$ measurement for all pulsars. The injected true values (\textcolor{red}{red solid line}) and priors (\textcolor{blue}{blue dashed line}) are also shown.}
    \label{fig:700_improved}
\end{figure*}

\section{Pulsar data}
Table~\ref{tab:pdd_pulsar_parameters} includes the data used for $\ddot{P}$ search, and Table~\ref{tab:pbdot_pulsar_parameters} includes the data used for $\dot{P}_b$ search. By the criterion described in Sec.~\ref{sec:dataset}, the 15 pulsars in bold font in Table~\ref{tab:pbdot_pulsar_parameters} are used in setting the upper limits.

\label{sec:pulsar data}

\renewcommand{\arraystretch}{1.5}

\begin{table*}[h]
\centering
\scriptsize
\begin{tabular}{lrrrrrrr}
\hline
         Pulsar &     $l$ (deg) &     $b$ (deg) &     d (kpc) &    $\Delta\dot{P}_b/P_b(10^{-18}s^{-1})$ \\
\hline
     B1259-63 & 304.184 &  -0.992 &      2.6(4) &      130(70) \\
     \textbf{B1534+12} &  19.848 &  48.341 &     0.94(7) &      0.2(1) \\
     B1913+16 &  49.968 &   2.122 &     
     4(1) &      -1.26(8) \\
   J0348+0432 & 183.337 & -36.774 &        1.6(9) &      -2(6) \\
   \textbf{J0437-4715} & 253.394 & -41.963 &     0.16(1) &     0.05(6) \\
   \textbf{J0613-0200} & 210.413 &  -9.305 &       1.11(5) &       -0.1(2) \\
\textbf{J0737-3039A/B} & 245.236 &  -4.505 &     0.74(6) &    -0.5(40) \\
   J0740+6620 & 149.730 &  29.599 &      1.2(2) &       -0.04(60) \\
   J0751+1807 & 202.730 &  21.086 &      1.4(4) &     -0.3(10) \\
   \textbf{J1012+5307} & 160.347 &  50.858 &     0.85(1) &      0.2(2) \\
   J1017-7156 & 291.558 & -12.553 &       
   1(2) &      0.5(6) \\
   \textbf{J1022+1001} & 231.795 &  51.101 &     0.85(6) &     -0.08(11) \\
   J1125-6014 & 292.504 &   0.894 &        1.5(8) &     -0.02(60) \\
   \textbf{J1455-3330} & 330.722 &  22.562 &     0.76(6) &      0.6(3) \\
   \textbf{J1537+1155} &  19.850 &  48.340 &      1.2(2) &       -0.09(40) \\
   J1600-3053 & 344.090 &  16.451 &       1.39(4) &       0.18(8) \\
   \textbf{J1603-7202} & 316.630 & -14.496 &      0.5(3) &       0.29(8) \\
   J1614-2230 & 352.636 &  20.192 &     0.69(5) &      0.3(2) \\
   \textbf{J1640+2224} &  41.051 &  38.271 &      1.0(2) &      0.4(2) \\
   \textbf{J1713+0747} &  28.751 &  25.223 &       1.31(4) &      -0.03(2) \\
   J1738+0333 &  27.721 &  17.742 &      1.7(1) &      0.03(30) \\
   J1741+1351 &  37.885 &  21.641 &        1.8(4) &        23(6) \\
   \textbf{J1909-3744} & 359.731 & -19.596 &       1.14(1) &       0.05(3) \\
   J1933-6211 & 334.431 & -28.632 &        1.2(4) &     -0.3(1) \\
   J2043+1711 &  61.919 & -15.313 &        1.6(2) &    107(60) \\
   J2129-5721 & 338.005 & -43.570 &        0.6(4) &        2.5(2) \\
   \textbf{J2145-0750} &  47.777 & -42.084 &       0.62(2) &         0.11(5) \\
   \textbf{J2222-0137} &  62.018 & -46.075 & 0.2681(1) & 0.00(5) \\
   \textbf{J2234+0611} &  72.991 & -43.006 &       0.95(4) &       -0.4(9) \\
   J2339-0533 &  81.349 & -62.476 &      1.4(3) &       8(5) \\
\hline
\end{tabular}
\caption{Complete table of pulsar data including Galactic coordinates, distance, and fractional acceleration in orbital period $\Delta\dot{P}_b/P_b$. This quantity includes all contributions that should sum up to $a_{\text{GW}}$ as in Eq.~\ref{eq:Pbdot_measurement}. Uncertainties are reported as 1 standard deviation. The contribution from Galactic acceleration is calculated using galpy~\cite{Bovy2015}, and all distances are measured using parallax. The distance and timing parameter data can be found in~\cite{Moran:2023, DeRoccoDror2024}. The pulsars satisfying the distance measurement criterion discussed in Sec.~\ref{sec:dataset} are in bold face.}
\label{tab:pbdot_pulsar_parameters}
\end{table*}

\renewcommand{\arraystretch}{1.5}

\begin{table*}[b]
\centering
\scriptsize
\begin{tabular}{lrrrrrr}
\hline
Pulsar & $l$ (deg) & $b$ (deg) & $d$ (kpc) & $\ddot{P}_{\text{obs}}/P$ ($10^{-30}\,\mathrm{s}^{-2}$) & $\sigma_{\text{RN}}$ ($10^{-30}\,\mathrm{s}^{-2}$) & Ref. \\
\hline
J0030+0451 & 113.141 & -57.611 & 0.3(1)\cite{Arzoumanian_2018} & $-4(4)$ & 0.54 & \cite{Desvignes2016} \\
J0034-0534 & 111.492 & -68.069 & 1.3(5)\cite{yao2017new} & $0(20)$ & 0 & \cite{Desvignes2016} \\
J0218+4232 & 139.508 & -17.527 & 3(1)\cite{Verbiest_2014} & $-2(5)$ & 0.14 & \cite{Desvignes2016} \\
J0437-4715 & 253.394 & -41.963 & 0.16(1)\cite{Moran:2023} & $-1(1)$ & 0 & \cite{Reardon2016} \\
J0610-2100 & 227.747 & -18.184 & 3(1)\cite{yao2017new} & $0(50)$ & 0 & \cite{Desvignes2016} \\
J0613-0200 & 210.413 & -9.305 & 1.11(5)\cite{Moran:2023} & $0.6(6)$ & 0.13 & \cite{Desvignes2016} \\
J0621+1002 & 200.570 & -2.013 & 0.4(2)\cite{yao2017new} & $-70(30)$ & 1.28 & \cite{Desvignes2016} \\
J0711-6830 & 279.531 & -23.280 & 0.11(4)\cite{yao2017new} & $1(1)$ & 0 & \cite{Reardon2016} \\
J0751+1807 & 202.730 & 21.086 & 1.4(4)\cite{Moran:2023} & $0(2)$ & 0 & \cite{Desvignes2016} \\
J0900-3144 & 256.162 & 9.486 & 0.8(4)\cite{Desvignes2016} & $-10(20)$ & 0 & \cite{Desvignes2016} \\
J1012+5307 & 160.347 & 50.858 & 0.85(1)\cite{Moran:2023} & $0.4(7)$ & 0.01 & \cite{Desvignes2016} \\
J1022+1001 & 231.795 & 51.101 & 0.85(6)\cite{Moran:2023} & $-2(1)$ & 0.01 & \cite{Desvignes2016} \\
J1045-4509 & 280.851 & 12.254 & 0.3(2)\cite{Bobakov_2019} & $-2(7)$ & 0.05 & \cite{Reardon2016} \\
J1455-3330 & 330.722 & 22.562 & 0.76(6)\cite{Moran:2023} & $6(20)$ & 0.17 & \cite{Desvignes2016} \\
J1600-3053 & 344.090 & 16.451 & 1.39(4)\cite{Moran:2023} & $4(5)$ & 0.06 & \cite{Reardon2016} \\
J1603-7202 & 316.630 & -14.496 & 0.5(3)\cite{Moran:2023} & $1(4)$ & 0.02 & \cite{Reardon2016} \\
J1640+2224 & 41.051 & 38.271 & 1.0(2)\cite{Moran:2023} & $-0.9(9)$ & 0 & \cite{Desvignes2016} \\
J1643-1224 & 5.669 & 21.218 & 0.8(2)\cite{Desvignes2016} & $-2(2)$ & 0.01 & \cite{Desvignes2016} \\
J1713+0747 & 28.751 & 25.223 & 1.31(4)\cite{Moran:2023} & $-0.5(5)$ & 0.38 & \cite{Desvignes2016} \\
J1721-2457 & 8.267 & 6.751 & 1.4(6)\cite{yao2017new} & $-30(70)$ & 0.01 & \cite{Desvignes2016} \\
J1730-2304 & 3.137 & 6.023 & 0.44(5)\cite{Shamohammadi:2024uvg} & $0(2)$ & 0 & \cite{Reardon2016} \\
J1732-5049 & 340.029 & -9.454 & 1.9(7)\cite{yao2017new} & $20(20)$ & 0.03 & \cite{Reardon2016} \\
J1738+0333 & 27.721 & 17.742 & 1.7(1)\cite{yao2017new} & $-30(90)$ & 0 & \cite{Desvignes2016} \\
J1744-1134 & 14.794 & 9.180 & 0.38(1)\cite{Shamohammadi:2024uvg} & $0.8(8)$ & 0.02 & \cite{Desvignes2016} \\
J1751-2857 & 0.646 & -1.124 & 1.1(4)\cite{yao2017new} & $-10(50)$ & 0 & \cite{Desvignes2016} \\
J1801-1417 & 14.546 & 4.162 & 2(1)\cite{Shamohammadi:2024uvg} & $-30(100)$ & 0.02 & \cite{Desvignes2016} \\
J1802-2124 & 8.382 & 0.611 & 0.6(4)\cite{Desvignes2016} & $10(60)$ & 0.01 & \cite{Desvignes2016} \\
J1804-2717 & 3.508 & -2.736 & 0.8(3)\cite{yao2017new} & $-40(40)$ & 0 & \cite{Desvignes2016} \\
J1843-1113 & 22.055 & -3.397 & 5(3)\cite{gaia_2023} & $-7(20)$ & 0.05 & \cite{Desvignes2016} \\
J1853+1303 & 44.875 & 5.367 & 0.9(4)\cite{Desvignes2016} & $-30(20)$ & 0 & \cite{Desvignes2016} \\
B1855+09 & 42.290 & 3.060 & 0.9(2)\cite{Verbiest_2012} & $1(2)$ & 0.03 & \cite{Desvignes2016} \\
J1909-3744 & 359.731 & -19.596 & 1.14(1)\cite{Moran:2023} & $0.6(9)$ & 0.02 & \cite{Desvignes2016} \\
J1910+1256 & 46.564 & -1.795 & 2.5(5)\cite{K_t_kc__2022} & $30(20)$ & 0 & \cite{Desvignes2016} \\
J1911+1347 & 25.137 & -9.579 & 2.0(3)\cite{K_t_kc__2022} & $14(8)$ & 0 & \cite{Desvignes2016} \\
J1911-1114 & 47.518 & 1.809 & 1.1(4)\cite{yao2017new} & $20(50)$ & 0 & \cite{Desvignes2016} \\
J1918-0642 & 30.027 & -9.123 & 1.2(3)\cite{Shamohammadi:2024uvg} & $0(8)$ & 2.46 & \cite{Desvignes2016} \\
B1953+29 & 65.839 & 0.443 & 6(3)\cite{yao2017new} & $-20(50)$ & 0 & \cite{Desvignes2016} \\
J2010-1323 & 29.446 & -23.540 & 3(2)\cite{Shamohammadi:2024uvg} & $20(20)$ & 0 & \cite{Desvignes2016} \\
J2019+2425 & 64.746 & -6.624 & 1.2(5)\cite{yao2017new} & $-500(900)$ & 0 & \cite{Desvignes2016} \\
J2033+1734 & 60.857 & -13.154 & 1.7(7)\cite{yao2017new} & $40(100)$ & 0 & \cite{Desvignes2016} \\
J2124-3358 & 10.925 & -45.438 & 0.48(4)\cite{Shamohammadi:2024uvg} & $0(3)$ & 0.02 & \cite{Reardon2016} \\
J2129-5721 & 338.005 & -43.570 & 0.6(4)\cite{Moran:2023} & $-1(2)$ & 0 & \cite{Reardon2016} \\
J2145-0750 & 47.777 & -42.084 & 0.62(2)\cite{Moran:2023} & $-2(1)$ & 0.28 & \cite{Desvignes2016} \\
J2229+2643 & 87.693 & -26.284 & 1.8(7)\cite{yao2017new} & $-20(20)$ & 0 & \cite{Desvignes2016} \\
J2317+1439 & 91.361 & -42.360 & 2.0(4)\cite{Arzoumanian_2018} & $-1(3)$ & 0 & \cite{Desvignes2016} \\
J2322+2057 & 96.515 & -37.310 & 0.8(2)\cite{Shamohammadi:2024uvg} & $30(70)$ & 0 & \cite{Desvignes2016} \\
\hline
\end{tabular}
\caption{Pulsar timing parameters including Galactic coordinates, distance, fractional spin-down acceleration $\ddot{P}_{\text{obs}}/P$ as in Eq.~\ref{eq:pddcalc}, variance induced by red noise $\sigma_{\text{RN}}$, and references. The uncertainties are reported as 1 standard deviation. For pulsars without a parallax measurement, the electron density model YMW 17~\cite{yao2017new} is used to estimate the distance. A relative uncertainty of $40\%$ is assumed in such cases. References for both the pulsar distances and the timing parameter data are included.}
\label{tab:pdd_pulsar_parameters}
\end{table*}

\clearpage
\twocolumngrid

\bibliographystyle{apsrev4-2}
\bibliography{bib}

\end{document}